\documentclass[11pt,english]{article}
\usepackage{lmodern}
\usepackage[T1]{fontenc}
\usepackage[latin9]{inputenc}
\usepackage{xcolor}
\usepackage{amsmath}
\usepackage{amssymb}
\usepackage{cancel}
\usepackage{stackrel}
\usepackage{graphicx}
\usepackage{setspace}
\usepackage{wasysym}
\usepackage{colortbl}
\usepackage{multirow}
\makeatletter

\usepackage{jheppub}

\usepackage{longtable}
\usepackage[font=small,labelfont=sc]{caption}
\usepackage[]{subcaption}
\usepackage{float}
\usepackage[compat=1.1.0]{tikz-feynman}
\tikzfeynmanset{warn luatex=false}
\usepackage[normalem]{ulem}
\usepackage{geometry}
\usepackage{booktabs}
\hypersetup{
linkcolor=purple,
urlcolor=purple,
citecolor=purple,
linktocpage=true
}

\definecolor{darkred}{rgb}{0.6,0,0}
\definecolor{darkpurple}{rgb}{0.5,0,0.5}

\def\z2{$\mathbb{Z}_2$}
\def\z3{$\mathbb{Z}_3$}
\def\321{$SU(3)_c \times SU(2)_L \times U(1)_Y$}

\def\555{\ensuremath{SU(5)^3}}

\def\24{\ensuremath{\mathbf{24}}}

\def\SU15#1{\ensuremath{\mathbf{#1}}}

\definecolor{ForestGreen}{RGB}{34,139,34}

\providecommand{\tabularnewline}{\\}

\title{\boldmath Minimal complete tri-hypercharge theories of flavour}

\author[a]{Mario Fern\'andez Navarro,}
\author[b]{Stephen F. King}
\author[c,d]{and Avelino Vicente}
\affiliation[a]{School of Physics \& Astronomy, University of Glasgow, Glasgow G12 8QQ, UK}
\affiliation[b]{School of Physics \& Astronomy, University of Southampton, Southampton SO17 1BJ, UK}
\affiliation[c]{Instituto de F\'isica Corpuscular, CSIC-Universitat de Val\`encia, 46980 Paterna, Spain}
\affiliation[d]{Departament de F\'isica Te\`orica, Universitat de Val\`encia, 46100 Burjassot, Spain}

\emailAdd{Mario.FernandezNavarro@glasgow.ac.uk}
\emailAdd{S.F.King@soton.ac.uk}
\emailAdd{avelino.vicente@ific.uv.es}

\abstract{
The tri-hypercharge proposal introduces a separate gauged weak hypercharge assigned to each fermion family as the origin of flavour. This is arguably one of the simplest setups for building ``gauge non-universal theories of flavour'' or ``flavour deconstructed theories''. 
In this paper we propose and study two minimal but ultraviolet complete and renormalisable tri-hypercharge models. 
We show that both models, which differ only by the heavy messengers that complete the effective theory, are able to explain the observed patterns of fermion masses and mixings (including neutrinos) with all fundamental coefficients being of $\mathcal{O}(1)$. In fact, both models translate the complicated flavour structure of the Standard Model into three simple physical scales above electroweak symmetry breaking, completely correlated with each other, that carry meaningful phenomenology. In particular, the heavy messenger sector determines the origin and size of fermion mixing, which controls the size and nature of the flavour-violating currents mediated by the two heavy $Z'$ gauge bosons of the theory. The phenomenological implications of the two minimal models are compared. In both models the lightest $Z'$ remains discoverable in dilepton searches at the LHC Run 3.
}

\makeatother

\usepackage{babel}
\begin{document}
\makeatletter
\gdef\@fpheader{}
\makeatother

\maketitle \flushbottom
\raggedbottom

\clearpage
\pagenumbering{arabic}
\setcounter{page}{1}
\allowdisplaybreaks

\section{Introduction}
The Standard Model (SM) of particle physics, though highly successful, offers no insight into the origin of the three fermion families nor the pattern of fermion Yukawa couplings, which results in a hierarchical mass spectrum for quarks and charged leptons that may be expressed at low energies as~\cite{PDG:2022ynf}
\begin{alignat}{2}
 & m_{t}\sim\frac{v_{\mathrm{SM}}}{\sqrt{2}}\,,\qquad & m_{c}\sim\lambda^{3.3}\frac{v_{\mathrm{SM}}}{\sqrt{2}}\,,\qquad & m_{u}\sim\lambda^{7.5}\frac{v_{\mathrm{SM}}}{\sqrt{2}}\,,\\
 & m_{b}\sim\lambda^{2.5}\frac{v_{\mathrm{SM}}}{\sqrt{2}}\,,\qquad & m_{s}\sim\lambda^{5.0}\frac{v_{\mathrm{SM}}}{\sqrt{2}}\,,\qquad & m_{d}\sim\lambda^{7.0}\frac{v_{\mathrm{SM}}}{\sqrt{2}}\,,\\
 & m_{\tau}\sim\lambda^{3.0}\frac{v_{\mathrm{SM}}}{\sqrt{2}}\,,\qquad & m_{\mu}\sim\lambda^{4.9}\frac{v_{\mathrm{SM}}}{\sqrt{2}}\,,\qquad & m_{e}\sim\lambda^{8.4}\frac{v_{\mathrm{SM}}}{\sqrt{2}}\,,
\end{alignat}
where $v_{\mathrm{SM}}\simeq246\,\mathrm{GeV}$ is the SM vacuum expectation
value (VEV) and $\lambda\simeq0.224$ is the Wolfenstein parameter
which parameterises the CKM matrix as
\begin{alignat}{2}
 & V_{us}\sim\lambda\,,\qquad & V_{cb}\sim\lambda^{2}\,,\qquad & V_{ub}\sim\lambda^{3}\,.
\end{alignat}
In addition, the discovery of very small neutrino masses and large PMNS mixing \cite{deSalas:2020pgw,Gonzalez-Garcia:2021dve},
\begin{alignat}{2}
 & \tan\theta_{23}\sim1\,,\qquad & \tan\theta_{12}\sim\frac{1}{\sqrt{2}}\,,\qquad & \theta_{13}\sim\frac{\lambda}{\sqrt{2}}\,,
\end{alignat}
has made the flavour puzzle difficult to ignore.

A tentative approach to explain the flavour puzzle consists of embedding the SM in a larger gauge symmetry that contains a separate gauge group for each fermion family, with the light Higgs doublet(s) originating from the third family group. This general idea is sometimes denoted in the literature as either ``gauge non-universal theories of flavour'' or ``flavour deconstructed theories''. Different possibilities have been discussed since the early 80s~\cite{Salam:1979p,Rajpoot:1980ib,Georgi:1981gj,Li:1981nk,Ma:1987ds,Ma:1988dn,Li:1992fi,Hill:1994hp,Barbieri:1994cx,Carone:1995ge,Muller:1996dj,Malkawi:1996fs,Dvali:2000ha,Asaka:2004ry,Babu:2007mb}, with new ideas appearing also in recent times~\cite{Craig:2011yk,Panico:2016ull,Bordone:2017bld,Greljo:2018tuh,Fuentes-Martin:2020bnh,Allwicher:2020esa,Fuentes-Martin:2020pww,Fuentes-Martin:2022xnb,Davighi:2022fer,Davighi:2022bqf,Davighi:2023iks,FernandezNavarro:2023rhv,Davighi:2023evx,FernandezNavarro:2023hrf,Davighi:2023xqn,Capdevila:2024gki,Barbieri:2023qpf,Fuentes-Martin:2024fpx}. Remarkably, these theories distinguish the three fermion families (hence avoiding family replication) and are able to explain the flavour structure of the SM with all fundamental couplings being of $\mathcal{O}(1)$. Out of them, arguably the simplest one involves just assigning one separate gauged weak hypercharge to each fermion family \cite{FernandezNavarro:2023rhv}
\begin{equation}
SU(3)_{c}\times SU(2)_{L}\times U(1)_{Y_{1}}\times U(1)_{Y_{2}}\times U(1)_{Y_{3}}\, .
\label{TH}
\end{equation}
Under the extended gauge group 
in Eq.~\eqref{TH} the $i$-th fermion family only carries $Y_{i}$
hypercharge, with the other hypercharges set equal to zero (see Table~\ref{tab:Field_content}),
where $Y=Y_{1}+Y_{2}+Y_{3}$ is equal to the usual SM weak gauged hypercharge.
Anomalies cancel separately for each family, as in the SM, but without family
replication. 
Two Higgs doublets are introduced which carry third family hypercharge, $H^{u,d}_{3}\sim(\mathbf{1,2})_{(0,0,\pm1/2)}$, leading to renormalisable Yukawa couplings only involving the third family. 
Yukawa couplings involving the first and second families are generated only after the gauge symmetry is broken to the SM gauge group, leading to a hierarchical pattern of Yukawa couplings.

The so-called ``tri-hypercharge'' (TH) gauge group 
in Eq.~\eqref{TH} is broken down to the SM gauge group by the VEVs
of appropriate scalars that carry family hypercharges which add up
to zero. The breaking chain 
\begin{flalign}
 & SU(3)_{c}\times SU(2)_{L}\times U(1)_{Y_{1}}\times U(1)_{Y_{2}}\times U(1)_{Y_{3}}\label{eq:Model1}\\
 & \xrightarrow{v_{12}} SU(3)_{c}\times SU(2)_{L}\times U(1)_{Y_{1}+Y_{2}}\times U(1)_{Y_{3}}\\
 & \xrightarrow{v_{23}} SU(3)_{c}\times SU(2)_{L}\times U(1)_{Y_{1}+Y_{2}+Y_{3}}\,.\label{eq:Model3}
\end{flalign}
leads to a successful generation of charged fermion mass hierarchies
and quark mixing. There generally exists a hierarchy between $v_{12}$
and $v_{23}$ that is related to the origin of the fermion mass hierarchies, as we shall see.
Typically $v_{23}\sim\mathcal{O}(1-10\,\mathrm{TeV})$ and $v_{12}\sim\mathcal{O}(100-10^{4}\,\mathrm{TeV})$
are allowed by current data.

In \cite{FernandezNavarro:2023rhv} two of us provided a general recipe for model building in the tri-hypercharge framework based on a spurion formalism, we proposed simple models in an effective field theory (EFT) framework (without specifying the heavy messengers that complete the theory) and studied the main phenomenological predictions of the setup, which are connected to heavy gauge bosons that arise after the spontaneous breaking of the TH group. Then, in \cite{FernandezNavarro:2023hrf} the three of us built an  ultraviolet (UV) complete tri-hypercharge theory that arises from a gauge unified framework based on assigning a separate $SU(5)$ gauge group to each fermion family, with the three $SU(5)$s related by a cyclic $\mathbb{Z}_{3}$ permutation symmetry that enforces a single gauge coupling at the grand unification (GUT) scale.

In this paper we provide two examples of minimal but ultraviolet complete tri-hypercharge models that contain the minimal amount of total degrees of freedom and representations, being amongst the most minimal models in the literature \cite{Salam:1979p,Rajpoot:1980ib,Georgi:1981gj,Li:1981nk,Ma:1987ds,Ma:1988dn,Li:1992fi,Hill:1994hp,Barbieri:1994cx,Carone:1995ge,Muller:1996dj,Malkawi:1996fs,Dvali:2000ha,Asaka:2004ry,Babu:2007mb,Craig:2011yk,Panico:2016ull,Bordone:2017bld,Greljo:2018tuh,Fuentes-Martin:2020bnh,Allwicher:2020esa,Fuentes-Martin:2020pww,Fuentes-Martin:2022xnb,Davighi:2022fer,Davighi:2022bqf,Davighi:2023iks,FernandezNavarro:2023rhv,Davighi:2023evx,FernandezNavarro:2023hrf,Davighi:2023xqn,Capdevila:2024gki,Barbieri:2023qpf,Fuentes-Martin:2024fpx}. In particular, the rank of the tri-hypercharge symmetry is the most minimal. We show that both models, which differ only by the heavy messengers that complete the effective theory, are able to explain the observed patterns of fermion masses and mixings (including neutrinos) with all fundamental coefficients being of $\mathcal{O}(1)$.

In fact, we shall see that the two minimal models presented here translate the complicated flavour structure of the Standard Model into three simple physical scales above electroweak symmetry breaking, completely correlated with each other and with the flavour hierarchies of the SM, that carry meaningful phenomenology. In this regard, we find that the heavy messengers crucially dictate the origin and size of fermion mixing, which controls the size and nature of the flavour-violating currents mediated by the two heavy $Z'$ gauge bosons of the theory. The phenomenological implications of the two minimal models are compared, where the lightest $Z'$ is discoverable in dilepton searches at the LHC Run 3 in both cases.

The layout of the paper is as follows. In Section~\ref{sec:MinimalModels} we first introduce the tri-hypercharge theory from a bottom-up perspective and an EFT point of view, and then we show two minimal UV-complete models. In the model of Section~\ref{sec:Model1} all the heavy messengers that complete the effective theory are vector-like fermions (hence having the simplest scalar sector and potential), whereas the model of Section~\ref{sec:Model2} offers a combination of heavy Higgs doublets and vector-like fermions that contains the minimal amount of degrees of freedom and representations. In Section~\ref{sec:Neutrinos} we discuss the neutrino sector that is common for both complete models. Then in Section~\ref{sec:Phenomenology} we discuss the model-dependent phenomenology associated to the two complete models, which is mostly dominated by flavour-changing neutral currents (FCNCs). Finally, Section~\ref{sec:Conclusions} concludes the paper. Additional technical details are given in three separate appendices.

\begin{table}[t]
\begin{centering}
\begin{tabular}{lcccc}
\toprule 
Field  & $U(1)_{Y_{1}}$  & $U(1)_{Y_{2}}$  & $U(1)_{Y_{3}}$ & $SU(3)_{c}\times SU(2)_{L}$\tabularnewline
\midrule 
$q_{1}$  & $1/6$  & 0  & 0 & $(\mathbf{3,2})$\tabularnewline
$u_{1}^{c}$  & $-2/3$  & 0  & 0 & $(\mathbf{\overline{3},1})$\tabularnewline
$d_{1}^{c}$  & $1/3$  & 0  & 0 & $(\mathbf{\overline{3},1})$\tabularnewline
$\ell_{1}$  & $-1/2$  & 0  & 0 & $(\mathbf{1,2})$\tabularnewline
$e_{1}^{c}$  & $1$  & 0  & 0 & $(\mathbf{1,1})$\tabularnewline
\midrule 
$q_{2}$  & 0  & $1/6$  & 0 & $(\mathbf{3,2})$\tabularnewline
$u_{2}^{c}$  & 0  & $-2/3$  & 0 & $(\mathbf{\overline{3},1})$\tabularnewline
$d_{2}^{c}$  & 0  & $1/3$  & 0 & $(\mathbf{\overline{3},1})$\tabularnewline
$\ell_{2}$  & 0  & $-1/2$  & 0 & $(\mathbf{1,2})$\tabularnewline
$e_{2}^{c}$  & 0  & $1$  & 0 & $(\mathbf{1,1})$\tabularnewline
\midrule 
$q_{3}$  & 0  & 0  & $1/6$ & $(\mathbf{3,2})$\tabularnewline
$u_{3}^{c}$  & 0  & 0  & $-2/3$ & $(\mathbf{\overline{3},1})$\tabularnewline
$d_{3}^{c}$  & 0  & 0  & $1/3$ & $(\mathbf{\overline{3},1})$\tabularnewline
$\ell_{3}$  & 0  & 0  & $-1/2$ & $(\mathbf{1,2})$\tabularnewline
$e_{3}^{c}$  & 0  & 0  & $1$ & $(\mathbf{1,1})$\tabularnewline
\midrule 
$\nu_{1}^{c}$  & 0  & 0  & 0 & $(\mathbf{1,1})$\tabularnewline
$\nu_{2}^{c}$  & 0  & 0  & 0 & $(\mathbf{1,1})$\tabularnewline
\midrule 
$H_{3}^{u}$  & 0  & 0  & $1/2$ & $(\mathbf{1,2})$\tabularnewline
$H_{3}^{d}$  & 0  & 0  & $-1/2$ & $(\mathbf{1,2})$\tabularnewline
\midrule 
$\phi_{q12}$  & $-1/6$  & $1/6$  & 0 & $(\mathbf{1,1})$\tabularnewline
$\phi_{\ell12}$  & $1/2$  & $-1/2$  & 0 & $(\mathbf{1,1})$\tabularnewline
$\phi_{q23}$  & 0  & $-1/6$  & $1/6$ & $(\mathbf{1,1})$\tabularnewline
$\phi_{\ell23}$  & 0  & $1/2$  & $-1/2$ & $(\mathbf{1,1})$\tabularnewline
\bottomrule
\end{tabular}
\par\end{centering}
\caption[Caption for LOF]{Field content of the low-energy (EFT) tri-hypercharge theory. $q_{i}$
and $\ell_{i}$ (where $i=1,2,3$) are left-handed (LH) $SU(2)_{L}$
doublets of chiral quarks and leptons, while $u_{i}^{c}$, $d_{i}^{c}$
and $e_{i}^{c}$ are the CP conjugate\footnotemark right-handed (RH) quarks and
leptons (so that they become left-handed). Two singlet neutrinos $\nu_{1,2}^{c}$,
also taken as left-handed fields, will be relevant for the origin of neutrino
masses. $H_{3}^{u,d}$ are $SU(2)_{L}$ Higgs doublets, while $\phi_{q12,\ell12,q23,\ell23}$
are the so-called scalar \textit{hyperons} which spontaneously break
the three gauge hypercharges down to SM hypercharge (the diagonal
subgroup $Y=Y_{1}+Y_{2}+Y_{3}$). \label{tab:Field_content}}
\end{table}

\footnotetext{For a given fermion $\Psi$ that consists of two 4-component chiral spinors $\Psi_{L}=(\psi_{L},0)^{\mathrm{T}}$ and $\Psi_{R}=(0,\psi_{R})^{\mathrm{T}}$, where $\psi_{L}$ and $\psi_R$ are 2-component (Weyl) spinors, one may get rid of chiral indices by defining $\psi \equiv \psi_{L}$ and by taking the CP conjugate of $\psi_{R}$, i.e.~$\psi^{c}\equiv CP \psi_{R} (CP)^{-1}$, which is by construction a left-handed field. Then one can write the theory in terms of $\psi$ and $\psi^{c}$ only. For example, fermion bilinears among both notations are related by $\overline{\Psi}_{R}\Psi_L \rightarrow \psi^{c}\psi$. We also refer the reader to \cite{Dreiner:2008tw,Martin:2012us} for extensive reviews on 2-component spinor notation and techniques.}

\section{Minimal Models} \label{sec:MinimalModels}

We consider that at relatively high energies the gauge symmetry contains
one hypercharge abelian factor for each fermion family. These are
spontaneously broken via the VEVs of $\phi$ scalars (see Table~\ref{tab:Field_content})
carrying non-zero family hypercharges that add up to zero, so that
they are gauge singlets under the SM symmetry, and are denoted
as \textit{hyperons}, 
\begin{flalign}
 & SU(3)_{c}\times SU(2)_{L}\times U(1)_{Y_{1}}\times U(1)_{Y_{2}}\times U(1)_{Y_{3}}\\
 & \overset{\langle\phi_{q_{12}}\rangle,\left\langle \phi_{\ell12}\right\rangle }{\longrightarrow}SU(3)_{c}\times SU(2)_{L}\times U(1)_{Y_{1}+Y_{2}}\times U(1)_{Y_{3}}\\
 & \overset{\left\langle \phi_{q23}\right\rangle ,\left\langle \phi_{\ell23}\right\rangle }{\longrightarrow}SU(3)_{c}\times SU(2)_{L}\times U(1)_{Y_{1}+Y_{2}+Y_{3}}\,.\label{eq:SymmetryBreaking}
\end{flalign}
At renormalisable level, only third family Yukawa couplings are allowed.
Two (third family) Higgs doublets $H_{3}^{u,d}\sim(\mathbf{1,2})_{(0,0,\pm1/2)}$
perform spontaneous breaking of the electroweak symmetry, and explain
the mass hierarchies $m_{b,\tau}/m_{t}$ if we assume a type-II two
Higgs doublet model (2HDM) with $\tan\beta=\langle H_{3}^{u}\rangle/\langle H_{3}^{d}\rangle\sim\lambda^{-2}\approx20$.

The masses of the light charged fermions arise from non-renormalisable
operators involving the hyperon scalars. With the four\footnote{For a discussion about models with less than four hyperons, we refer the reader to Appendix~\ref{app:ModelsLessHyperons}.} hyperons of
Table~\ref{tab:Field_content}, one can generate all the spurions\footnote{In particular, we obtain the spurions $\eta\sim(-\frac{1}{6},-\frac{1}{3},\frac{1}{2})$ and $\hat{\eta}\sim(-\frac{1}{6},0,\frac{1}{6})$, needed to generate quark mixing, by combining the four hyperons of Table~\ref{tab:Field_content} as $\eta\sim\phi_{q12}\tilde{\phi}_{\ell23}$ and $\hat{\eta}\sim\phi_{q12}\phi_{q23}$.} introduced in Ref.~\cite{FernandezNavarro:2023rhv}, which populate the Yukawa matrices.
By working in an EFT framework, we can write\footnote{We note that in the 32 and 13 entries of the down-quark matrix, the
operators ${\displaystyle {\widetilde{\phi}}_{\ell23}{\widetilde{\phi}}_{q23}q_{3}d_{2}^{c}H_{3}^{d}}$
and ${\widetilde{\phi}}_{\ell12}{\widetilde{\phi}}_{q12}{\widetilde{\phi}}_{\ell23}{\widetilde{\phi}}_{q23}q_{3}d_{2}^{c}H_{3}^{d}$
contribute at similar order in the EFT expansion, but are not shown
for simplicity.} (ignoring $\mathcal{O}(1)$
dimensionless coefficients), 
\begin{flalign}
{\cal L} & =\begin{pmatrix}q_{1} & q_{2} & q_{3}\end{pmatrix}\begin{pmatrix}{\displaystyle \frac{\phi_{\ell12}}{\Lambda_{1}}\frac{{\phi}_{\ell23}}{\Lambda_{2}}} & {\displaystyle \frac{{\phi}_{q12}}{\Lambda_{1}}\frac{{\phi}_{\ell23}}{\Lambda_{2}}} & {\displaystyle \frac{{\phi}_{q12}}{\Lambda_{1}}\frac{{\phi}_{q23}}{\Lambda_{2}}}\\
\vspace{-0.14cm} & \vspace{-0.14cm} & \vspace{-0.14cm}\\
{\displaystyle \frac{\phi_{\ell12}}{\Lambda_{1}}\frac{{\widetilde{\phi}}_{q12}}{\Lambda_{1}}\frac{{\phi}_{\ell23}}{\Lambda_{2}}} & {\displaystyle \frac{{\phi}_{\ell23}}{\Lambda_{2}}} & {\displaystyle \frac{{\phi}_{q23}}{\Lambda_{2}}}\\
\vspace{-0.14cm} & \vspace{-0.14cm} & \vspace{-0.14cm}\\
{\displaystyle \frac{{\widetilde{\phi}}_{q12}}{\Lambda_{1}}\frac{\phi_{\ell12}}{\Lambda_{1}}\frac{{\phi}_{\ell23}}{\Lambda_{1}}}{\displaystyle \frac{{\widetilde{\phi}}_{q23}}{\Lambda_{2}}} & {\displaystyle \frac{{\phi}_{\ell23}}{\Lambda_{2}}}{\displaystyle \frac{{\widetilde{\phi}}_{q23}}{\Lambda_{2}}} & 1
\end{pmatrix}\begin{pmatrix}u_{1}^{c}\\
u_{2}^{c}\\
u_{3}^{c}
\end{pmatrix}H_{3}^{u}\label{eq:eff_Yukawa_up}\\
 & +\begin{pmatrix}q_{1} & q_{2} & q_{3}\end{pmatrix}\begin{pmatrix}{\displaystyle \frac{\tilde{\phi}_{\ell12}}{\Lambda_{1}}\frac{\tilde{\phi}_{\ell23}}{\Lambda_{2}}} & {\displaystyle \frac{\phi_{q12}}{\Lambda_{1}}\frac{\tilde{\phi}_{\ell23}}{\Lambda_{2}}} & {\displaystyle \frac{\phi_{q12}}{\Lambda_{1}}\frac{{\phi}_{q23}}{\Lambda_{2}}}\\
\vspace{-0.14cm} & \vspace{-0.14cm} & \vspace{-0.14cm}\\
{\displaystyle \frac{\tilde{\phi}_{\ell12}}{\Lambda_{1}}\frac{{\widetilde{\phi}}_{q12}}{\Lambda_{1}}\frac{\tilde{\phi}_{\ell23}}{\Lambda_{1}}} & {\displaystyle \frac{\tilde{\phi}_{\ell23}}{\Lambda_{2}}} & {\displaystyle \frac{{\phi}_{q23}}{\Lambda_{2}}}\\
\vspace{-0.14cm} & \vspace{-0.14cm} & \vspace{-0.14cm}\\
{\displaystyle \frac{\phi_{q12}^{2}}{\Lambda_{1}^{2}}\frac{{\phi}_{q23}^{2}}{\Lambda_{2}^{2}}} & {\displaystyle \frac{{\phi}_{q23}^{2}}{\Lambda_{2}^{2}}} & 1
\end{pmatrix}\begin{pmatrix}d_{1}^{c}\\
d_{2}^{c}\\
d_{3}^{c}
\end{pmatrix}H_{3}^{d}\label{eq:eff_Yukawa_down}\\
 & +\begin{pmatrix}\ell_{1} & \ell_{2} & \ell_{3}\end{pmatrix}\begin{pmatrix}{\displaystyle \frac{\tilde{\phi}_{\ell12}}{\Lambda_{1}}\frac{\tilde{\phi}_{\ell23}}{\Lambda_{2}}} & {\displaystyle \frac{\phi_{\ell12}}{\Lambda_{1}}\frac{\tilde{\phi}_{\ell23}}{\Lambda_{2}}} & {\displaystyle \frac{\phi_{\ell12}}{\Lambda_{1}}\frac{\phi_{\ell23}}{\Lambda_{2}}}\\
\vspace{-0.14cm} & \vspace{-0.14cm} & \vspace{-0.14cm}\\
{\displaystyle \frac{\tilde{\phi}_{\ell12}}{\Lambda_{1}}\frac{\tilde{\phi}_{\ell12}}{\Lambda_{1}}\frac{\tilde{\phi}_{\ell23}}{\Lambda_{2}}} & {\displaystyle \frac{\tilde{\phi}_{\ell23}}{\Lambda_{2}}} & {\displaystyle \frac{\phi_{\ell23}}{\Lambda_{2}}}\\
\vspace{-0.14cm} & \vspace{-0.14cm} & \vspace{-0.14cm}\\
{\displaystyle \frac{\tilde{\phi}_{\ell12}^{2}}{\Lambda_{1}^{2}}\frac{\tilde{\phi}_{\ell23}^{2}}{\Lambda_{2}^{2}}} & {\displaystyle \frac{\tilde{\phi}_{\ell23}^{2}}{\Lambda_{2}^{2}}} & 1
\end{pmatrix}\begin{pmatrix}e_{1}^{c}\\
e_{2}^{c}\\
e_{3}^{c}
\end{pmatrix}H_{3}^{d}\,+\mathrm{h.c.}\label{eq:eff_Yukawa_e}
\end{flalign}
Notice that the 23-hyperons generate the mass hierarchy $m_{2}/m_{3}$
along with naturally suppressed $V_{cb}$ and $V_{ub}$ CKM elements.
Then, the 12-hyperons discriminate the first family, generating the
mass hierarchy $m_{1}/m_{2}$, the Cabbibo angle $V_{us}$ and an
extra suppression for $V_{ub}$ with respect to $V_{cb}$, while the
origin of small neutrino masses and PMNS mixing is later discussed
in the complete model, see Section~\ref{sec:Neutrinos}.

In the following we UV-complete the above EFT by including heavy messengers
explicitly. Their masses provide the heavy scales $\Lambda_{1}$ and
$\Lambda_{2}$. Notice that we do not need to generate all the entries
in the Yukawa matrices of Eqs.~(\ref{eq:eff_Yukawa_up}-\ref{eq:eff_Yukawa_e})
to explain the flavour structure of the SM, and indeed some of them
will not be generated at tree-level in our complete models.

In the next two Sections we will show two minimal UV completions,
one where all the heavy messengers are vector-like (VL) fermions (hence
having the simplest scalar sector and potential), and another
one with a combination of heavy Higgs doublets and VL fermions
that contains the minimal amount of degrees of freedom and representations.

\subsection{Model 1}
\label{sec:Model1}

Our first proposal for a minimal, complete and renormalisable model
includes three vector-like $SU(2)_{L}$ singlet fermions for each
charged sector, as shown in Table~\ref{tab:UVmodel1}, that act as
heavy messengers of the effective Yukawa operators introduced before.
The full set of renormalisable Yukawa couplings and bare mass terms in the charged fermion
sector of the model is given by 
\begin{equation}
\mathcal{L}_{Y}^{u}=\left(\begin{array}{c|cccccc}
 & u_{1}^{c} & u_{2}^{c} & u_{3}^{c} & U_{12} & U_{13} & U_{23}\\
\hline q_{1} & 0 & 0 & 0 & 0 & y_{15}^{u}H_{3}^{u} & 0\\
q_{2} & 0 & 0 & 0 & 0 & 0 & y_{26}^{u}H_{3}^{u}\\
q_{3} & 0 & 0 & y_{3}^{u}H_{3}^{u} & 0 & 0 & 0\\
\overline{U}_{12} & y_{41}^{u}\phi_{\ell12} & y_{42}^{u}\phi_{q12} & 0 & M_{U_{12}} & y_{45}^{u}\tilde{\phi}_{\ell23} & 0\\
\overline{U}_{13} & 0 & 0 & 0 & y_{54}^{u}\phi_{\ell23} & M_{U_{13}} & y_{56}^{u}\phi_{q12}\\
\overline{U}_{23} & 0 & y_{62}^{u}\phi_{\ell23} & y_{63}^{u}\phi_{q23} & 0 & y_{65}^{u}\tilde{\phi}_{q12} & M_{U_{23}}
\end{array}\right)+\mathrm{h.c.}\label{eq:up_FullCouplings}
\end{equation}
\begin{equation}
\mathcal{L}_{Y}^{d}=\left(\begin{array}{c|cccccc}
 & d_{1}^{c} & d_{2}^{c} & d_{3}^{c} & D_{12} & D_{13} & D_{23}\\
\hline q_{1} & 0 & 0 & 0 & 0 & y_{15}^{d}H_{3}^{d} & 0\\
q_{2} & 0 & 0 & 0 & 0 & 0 & y_{26}^{d}H_{3}^{d}\\
q_{3} & 0 & 0 & y_{3}^{d}H_{3}^{d} & 0 & 0 & 0\\
\overline{D}_{12} & y_{41}^{d}\tilde{\phi}_{\ell12} & y_{42}^{d}\phi_{q12} & 0 & M_{D_{12}} & y_{45}^{d}\phi_{\ell23} & 0\\
\overline{D}_{13} & 0 & 0 & 0 & y_{54}^{d}\tilde{\phi}_{\ell23} & M_{D_{13}} & y_{56}^{d}\phi_{q12}\\
\overline{D}_{23} & 0 & y_{62}^{d}\tilde{\phi}_{\ell23} & y_{63}^{d}\phi_{q23} & 0 & y_{65}^{u}\tilde{\phi}_{q12} & M_{D_{23}}
\end{array}\right)+\mathrm{h.c.}\label{eq:down_FullCouplings}
\end{equation}
\begin{equation}
\mathcal{L}_{Y}^{e}=\left(\begin{array}{c|cccccc}
 & e_{1}^{c} & e_{2}^{c} & e_{3}^{c} & E_{12} & E_{13} & E_{23}\\
\hline \ell_{1} & 0 & 0 & 0 & 0 & y_{15}^{e}H_{3}^{d} & 0\\
\ell_{2} & 0 & 0 & 0 & 0 & 0 & y_{26}^{e}H_{3}^{d}\\
\ell_{3} & 0 & 0 & y_{3}^{e}H_{3}^{d} & 0 & 0 & 0\\
\overline{E}_{12} & y_{41}^{e}\tilde{\phi}_{\ell12} & y_{42}^{e}\phi_{\ell12} & 0 & M_{E_{12}} & y_{45}^{e}\phi_{\ell23} & 0\\
\overline{E}_{13} & 0 & 0 & 0 & y_{54}^{e}\tilde{\phi}_{\ell23} & M_{E_{13}} & y_{56}^{e}\phi_{\ell12}\\
\overline{E}_{23} & 0 & y_{62}^{e}\tilde{\phi}_{\ell23} & y_{63}^{e}\phi_{\ell23} & 0 & y_{65}^{e}\tilde{\phi}_{\ell12} & M_{E_{23}}
\end{array}\right)+\mathrm{h.c.}\label{eq:e_FullCouplings}
\end{equation}
\begin{table}
\begin{centering}
\begin{tabular}{ccccc}
\toprule 
Field  & $U(1)_{Y_{1}}$  & $U(1)_{Y_{2}}$  & $U(1)_{Y_{3}}$  & $SU(3)_{c}\times SU(2)_{L}$\tabularnewline
\midrule 
$H_{3}^{u,d}$  & 0  & 0  & $\pm1/2$  & $(\mathbf{1,2})$\tabularnewline
\midrule 
$\phi_{q_{12}}$  & -1/6  & 1/6  & 0  & $(\mathbf{1,1})$\tabularnewline
$\phi_{\ell_{12}}$  & 1/2  & -1/2  & 0  & $(\mathbf{1,1})$\tabularnewline
$\phi_{q_{23}}$  & 0  & -1/6  & 1/6  & $(\mathbf{1,1})$\tabularnewline
$\phi_{\ell_{23}}$  & 0  & 1/2  & -1/2  & $(\mathbf{1,1})$\tabularnewline
\midrule 
\rowcolor{yellow!10}$U_{12}$  & -1/6  & -1/2  & 0  & $(\mathbf{\overline{3},1})$\tabularnewline
\rowcolor{yellow!10}$U_{13}$  & -1/6  & 0  & -1/2  & $(\mathbf{\overline{3},1})$\tabularnewline
\rowcolor{yellow!10}$U_{23}$  & 0  & -1/6  & -1/2  & $(\mathbf{\overline{3},1})$\tabularnewline
\midrule 
\rowcolor{yellow!10}$D_{12}$  & -1/6  & 1/2  & 0  & $(\mathbf{\overline{3},1})$\tabularnewline
\rowcolor{yellow!10}$D_{13}$  & -1/6  & 0  & 1/2  & $(\mathbf{\overline{3},1})$\tabularnewline
\rowcolor{yellow!10}$D_{23}$  & 0  & -1/6  & 1/2  & $(\mathbf{\overline{3},1})$\tabularnewline
\midrule 
\rowcolor{yellow!10}$E_{12}$  & 1/2  & 1/2  & 0  & $(\mathbf{1,1})$\tabularnewline
\rowcolor{yellow!10}$E_{13}$  & 1/2  & 0  & 1/2  & $(\mathbf{1,1})$\tabularnewline
\rowcolor{yellow!10}$E_{23}$  & 0  & 1/2  & 1/2  & $(\mathbf{1,1})$\tabularnewline
\midrule
\rowcolor{yellow!10}$N_{12}$  & 1/2  & -1/2  & 0  & $(\mathbf{1,1})$\tabularnewline
\rowcolor{yellow!10}$N_{13}$  & 1/2  & 0  & -1/2  & $(\mathbf{1,1})$\tabularnewline
\rowcolor{yellow!10}$N_{23}$  & 0  & 1/2  & -1/2  & $(\mathbf{1,1})$\tabularnewline
\bottomrule
\end{tabular}
\par\end{centering}
\caption{Scalar and vector-like fermion content
of Model 1. Notice that for vector-like fermions (highlighted in yellow), only the left-handed fields $U,D,E,N$, 
with the same electric charges as the chiral fermions $u^c,d^c,e^c,\nu^c$ in Table~\ref{tab:Field_content}, are displayed, while their vector-like partners with conjugate quantum numbers are implicitly assumed to be present but not explicitly shown. \label{tab:UVmodel1}}
\end{table}

After integrating out the heavy VL charged fermions, the effective
Yukawa couplings for chiral charged fermions take the form 
\begin{flalign}
{\cal L} & =\begin{pmatrix}q_{1} & q_{2} & q_{3}\end{pmatrix}\begin{pmatrix}c_{11}^{u}{\displaystyle \frac{\phi_{\ell12}}{M_{U_{13}}}\frac{{\phi}_{\ell23}}{M_{U_{12}}}} & c_{12}^{u}{\displaystyle \frac{{\phi}_{q12}}{M_{U_{13}}}\frac{{\phi}_{\ell23}}{M_{U_{23}}}} & {\displaystyle c_{13}^{u}\frac{{\phi}_{q12}}{M_{U_{13}}}\frac{{\phi}_{q23}}{M_{U_{23}}}}\\
\vspace{-0.14cm} & \vspace{-0.14cm} & \vspace{-0.14cm}\\
c_{21}^{u}{\displaystyle \frac{\phi_{\ell12}}{M_{U_{12}}}}{\displaystyle \frac{\phi_{q12}}{M_{U_{13}}}}{\displaystyle \frac{\phi_{\ell23}}{M_{U_{23}}}} & c_{22}^{u}{\displaystyle \frac{{\phi}_{\ell23}}{M_{U_{23}}}} & c_{23}^{u}{\displaystyle \frac{{\phi}_{q23}}{M_{U_{23}}}}\\
\vspace{-0.14cm} & \vspace{-0.14cm} & \vspace{-0.14cm}\\
0 & 0 & c_{33}^{u}\label{eq:UpSectorYukawa}
\end{pmatrix}\begin{pmatrix}u_{1}^{c}\\
u_{2}^{c}\\
u_{3}^{c}
\end{pmatrix}H_{3}^{u}\\
 & +\begin{pmatrix}q_{1} & q_{2} & q_{3}\end{pmatrix}\begin{pmatrix}c_{11}^{d}{\displaystyle \frac{\tilde{\phi}_{\ell12}}{M_{D_{13}}}\frac{\tilde{\phi}_{\ell23}}{M_{D_{12}}}} & c_{12}^{d}{\displaystyle \frac{\phi_{q12}}{M_{D_{13}}}\frac{\tilde{\phi}_{\ell23}}{M_{D_{23}}}} & c_{13}^{d}{\displaystyle \frac{\phi_{q12}}{M_{D_{13}}}\frac{{\phi}_{q23}}{M_{D_{23}}}}\\
\vspace{-0.14cm} & \vspace{-0.14cm} & \vspace{-0.14cm}\\
c_{21}^{d}{\displaystyle \frac{\tilde{\phi}_{\ell12}}{M_{D_{12}}}}{\displaystyle \frac{\tilde{\phi}_{q12}}{M_{D_{13}}}}{\displaystyle \frac{\tilde{\phi}_{\ell23}}{M_{D_{23}}}} & c_{22}^{d}{\displaystyle \frac{\tilde{\phi}_{\ell23}}{M_{D_{23}}}} & c_{23}^{d}{\displaystyle \frac{{\phi}_{q23}}{M_{D_{23}}}}\\
\vspace{-0.14cm} & \vspace{-0.14cm} & \vspace{-0.14cm}\\
0 & 0 & c_{33}^{d}
\end{pmatrix}\begin{pmatrix}d_{1}^{c}\\
d_{2}^{c}\\
d_{3}^{c}
\end{pmatrix}H_{3}^{d}\\
 & +\begin{pmatrix}\ell_{1} & \ell_{2} & \ell_{3}\end{pmatrix}\begin{pmatrix}c_{11}^{e}{\displaystyle \frac{\tilde{\phi}_{\ell12}}{M_{E_{13}}}\frac{\tilde{\phi}_{\ell23}}{M_{E_{12}}}} & {\displaystyle {\displaystyle c_{12}^{e}\frac{\phi_{\ell12}}{M_{E_{13}}}\frac{\tilde{\phi}_{\ell23}}{M_{E_{23}}}}} & c_{13}^{e}{\displaystyle {\displaystyle \frac{\phi_{\ell12}}{M_{E_{13}}}\frac{{\phi}_{\ell23}}{M_{E_{23}}}}}\\
\vspace{-0.14cm} & \vspace{-0.14cm} & \vspace{-0.14cm}\\
c_{21}^{e}{\displaystyle \frac{\tilde{\phi}_{\ell12}}{M_{E_{12}}}}{\displaystyle \frac{\tilde{\phi}_{\ell12}}{M_{E_{13}}}}{\displaystyle \frac{\tilde{\phi}_{\ell23}}{M_{E_{23}}}} & c_{22}^{e}{\displaystyle \frac{\tilde{\phi}_{\ell23}}{M_{E_{23}}}} & {\displaystyle c_{23}^{e}\frac{\phi_{\ell23}}{M_{E_{23}}}}\\
\vspace{-0.14cm} & \vspace{-0.14cm} & \vspace{-0.14cm}\\
0 & 0 & c_{33}^{e}
\end{pmatrix}\begin{pmatrix}e_{1}^{c}\\
e_{2}^{c}\\
e_{3}^{c}
\end{pmatrix}H_{3}^{d}\,+\mathrm{h.c.}\,,\label{eq::ChargedLeptonsModel1}
\end{flalign}
where 
\begin{equation}
c^{u}=\begin{pmatrix}y_{15}^{u}y_{54}^{u}y_{41}^{u} & y_{15}^{u}y_{56}^{u}y_{62}^{u} & {\displaystyle y_{15}^{u}y_{56}^{u}y_{63}^{u}}\\
y_{26}^{u}y_{65}^{u}y_{54}^{u}y_{41}^{u} & y_{26}^{u}y_{62}^{u} & y_{26}^{u}y_{63}^{u}\\
0 & 0 & y_{3}^{u}
\end{pmatrix}\,,
\end{equation}
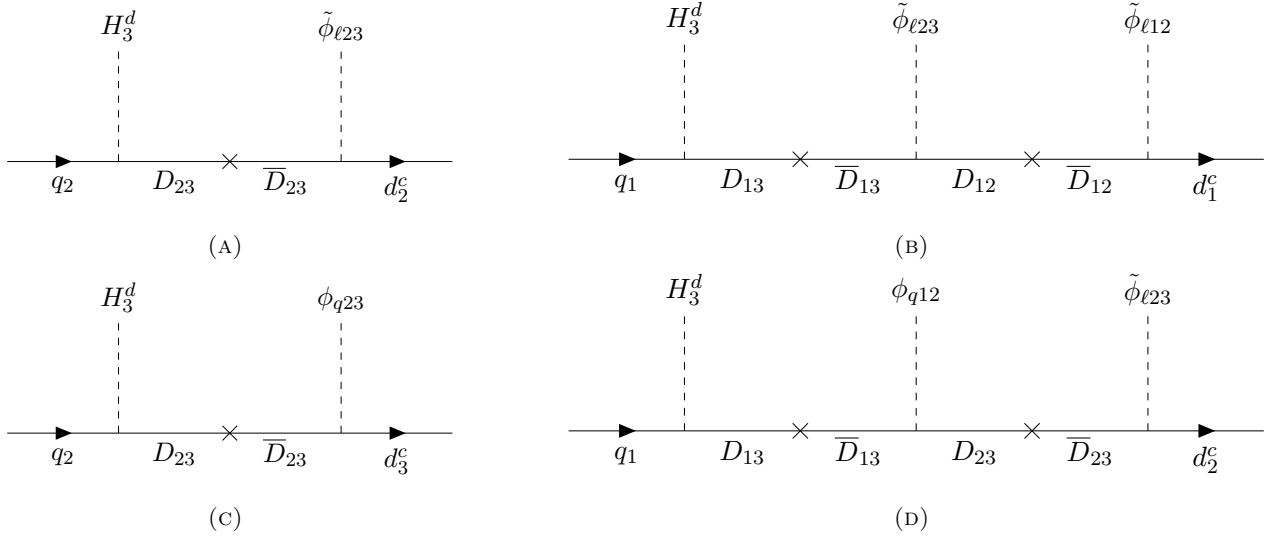
\begin{figure}[t]
\subfloat[]{\resizebox{.37\textwidth}{!}{ \begin{tikzpicture}
	\begin{feynman}
		\vertex (a);
		\vertex [right=16mm of a] (b);
		\vertex [right=16mm of b] (c);
		\vertex [right=16mm of c] (d);
		\vertex [right=16mm of d] (e);
		\vertex [above=16mm of b] (f1) {\(H^{d}_{3}\)};
		\vertex [above=16mm of d] (f2) {\(\tilde{\phi}_{\ell 23}\)};
		\diagram* {
			(a) -- [fermion, edge label'=\(q_{2}\), inner sep=6pt] (b) -- [scalar] (f1),
			(b) -- [edge label'=\(D_{23}\)] (c),
			(c) -- [edge label'=\(\overline{D}_{23}\),inner sep=2pt, insertion=0] (d) -- [scalar] (f2),
			(d) -- [fermion, edge label'=\(d_{2}^{c}\), inner sep=6pt] (e),
	};
	\end{feynman}
\end{tikzpicture} }

}$\qquad\quad$\subfloat[]{\resizebox{.57\textwidth}{!}{ \begin{tikzpicture}
	\begin{feynman}
		\vertex (a);
		\vertex [right=16mm of a] (b);
		\vertex [right=16mm of b] (c);
		\vertex [right=16mm of c] (d);
		\vertex [right=16mm of d] (e);
		\vertex [right=16mm of e] (f);
		\vertex [right=16mm of f] (g);
		\vertex [above=16mm of b] (f1) {\(H^{d}_{3}\)};
		\vertex [above=16mm of d] (f2) {\(\tilde{\phi}_{\ell 23}\)};
		\vertex [above=16mm of f] (f3) {\(\tilde{\phi}_{\ell 12}\)};
		\diagram* {
			(a) -- [fermion, edge label'=\(q_{1}\), inner sep=6pt] (b) -- [scalar] (f1),
			(b) -- [edge label'=\(D_{13}\)] (c),
			(c) -- [edge label'=\(\overline{D}_{13}\),inner sep=2pt, insertion=0] (d) -- [scalar] (f2),
			(d) -- [edge label'=\(D_{12}\)] (e),
			(e) -- [edge label'=\(\overline{D}_{12}\),inner sep=2pt, insertion=0] (f) -- [scalar] (f3),
			(f) -- [fermion, edge label'=\(d_{1}^{c}\), inner sep=6pt] (g),
	};
	\end{feynman}
\end{tikzpicture} }

}

\subfloat[]{\resizebox{.37\textwidth}{!}{ \begin{tikzpicture}
	\begin{feynman}
		\vertex (a);
		\vertex [right=16mm of a] (b);
		\vertex [right=16mm of b] (c);
		\vertex [right=16mm of c] (d);
		\vertex [right=16mm of d] (e);
		\vertex [above=16mm of b] (f1) {\(H^{d}_{3}\)};
		\vertex [above=16mm of d] (f2) {\(\phi_{q23}\)};
		\diagram* {
			(a) -- [fermion, edge label'=\(q_{2}\), inner sep=6pt] (b) -- [scalar] (f1),
			(b) -- [edge label'=\(D_{23}\)] (c),
			(c) -- [edge label'=\(\overline{D}_{23}\),inner sep=2pt, insertion=0] (d) -- [scalar] (f2),
			(d) -- [fermion, edge label'=\(d_{3}^{c}\), inner sep=6pt] (e),
	};
	\end{feynman}
\end{tikzpicture} }

}$\qquad\quad$\subfloat[]{\resizebox{.57\textwidth}{!}{ \begin{tikzpicture}
	\begin{feynman}
		\vertex (a);
		\vertex [right=16mm of a] (b);
		\vertex [right=16mm of b] (c);
		\vertex [right=16mm of c] (d);
		\vertex [right=16mm of d] (e);
		\vertex [right=16mm of e] (f);
		\vertex [right=16mm of f] (g);
		\vertex [above=16mm of b] (f1) {\(H^{d}_{3}\)};
		\vertex [above=16mm of d] (f2) {\(\phi_{q 12}\)};
		\vertex [above=16mm of f] (f3) {\(\tilde{\phi}_{\ell 23}\)};
		\diagram* {
			(a) -- [fermion, edge label'=\(q_{1}\), inner sep=6pt] (b) -- [scalar] (f1),
			(b) -- [edge label'=\(D_{13}\)] (c),
			(c) -- [edge label'=\(\overline{D}_{13}\),inner sep=2pt, insertion=0] (d) -- [scalar] (f2),
			(d) -- [edge label'=\(D_{23}\)] (e),
			(e) -- [edge label'=\(\overline{D}_{23}\),inner sep=2pt, insertion=0] (f) -- [scalar] (f3),
			(f) -- [fermion, edge label'=\(d_{2}^{c}\), inner sep=6pt] (g),
	};
	\end{feynman}
\end{tikzpicture} }

}\caption[]{Diagrams at the origin of light down-quark masses and mixing in Model 1. Similar diagrams are obtained in the
up-sector by replacing $d_{i}^{c}\rightarrow u_{i}^{c}$, $D_{12,13,23}\rightarrow U_{12,13,23}$,
$\tilde{\phi}_{\ell12,23}\rightarrow\phi_{\ell12,23}$ and $H_{3}^{d}\rightarrow H_{3}^{u}$,
and also for the charged lepton sector by replacing
$q_{i}\rightarrow\ell_{i}$, $d_{i}^{c}\rightarrow e_{i}^{c}$, $D_{12,13,23}\rightarrow E_{12,13,23}$
and $\phi_{q12,23}\rightarrow\phi_{\ell12,23}$.\label{fig:DiagramsModel1}}
\end{figure}and
similarly for the down and charged lepton sectors by replacing $u\rightarrow d,e$.
In terms of Feynman diagrams, the effective Yukawa couplings are generated
by mass insertions of the VL fermions as shown in Fig.~\ref{fig:DiagramsModel1},
with an extra (welcomed) mass insertion in all the diagrams involving
the first family. Once the hyperons develop VEVs, the effective Yukawa
matrices above generate naturally small Yukawa couplings for light
charged fermions. Now we shall fix the ratios of hyperon VEVs over
messenger masses in order to explain the flavour structure of the
SM.

Firstly, notice how the 23-messengers always appear together with
the 23-hyperons in the Yukawa matrices, and similarly the 12,13-messengers which distinguish the first family
always appear together with 12-hyperons. This suggests the presence
of one scale at which all the 23-messengers live - that we associate with
$\Lambda_{2}$ from the EFT formalism - and another, heavier scale where all
the 12,13-messengers live - that we associate with $\Lambda_{1}$.
The exception is the ratio $\phi_{\ell23}/M_{12}$ appearing in the
(1,1) entries, which provides a connection between \textit{both} scales.
This is a highly non-generic feature in this class of models, which
was \textit{not} anticipated by the EFT framework.

\begin{figure}[t]
\centering 
\includegraphics[scale=0.263]{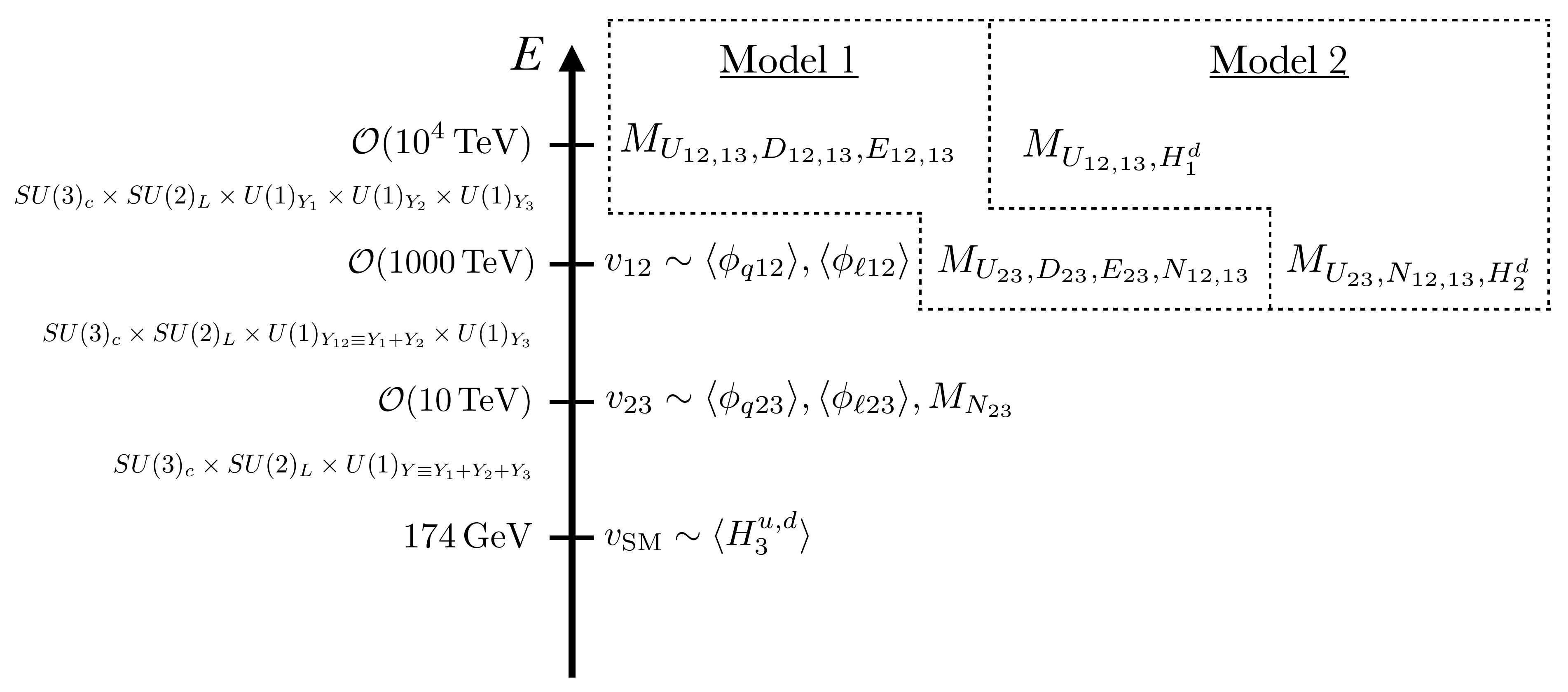} 
\caption{Illustrative spectrum of NP scales in Models 1 and 2. Notice that
the hyperons $\phi$, third family Higgs doublets $H^{u,d}_{3}$, messenger neutrinos $N$ and
messenger up-type quarks $U$ are common to both models. \label{fig:spectrum} }
\end{figure}

In fact, assuming for simplicity the two 23-VEVs to be degenerate and denoted generically as $\langle\phi_{23}\rangle$, the two 12-VEVs to be degenerate and denoted as $\langle\phi_{12}\rangle$,
the 23-messenger masses to be degenerate and denoted as $M_{23}$ and the 12,13-messenger
masses to be degenerate and denoted as $M_{12,13}$, then we find that the approximate
scaling
\begin{equation}
\langle\phi_{23}\rangle\ll M_{23}\sim\langle\phi_{12}\rangle\ll M_{12}\sim M_{13}
\end{equation}
provides an excellent explanation of the SM flavour structure, which
is translated to three simple and physical scales\footnote{In principle, the masses of VL fermions are arbitrary parameters and finding a dynamical mechanism that generates them at the scales of Fig.~\ref{fig:spectrum} is not straightforward. The VL fermions could be made chiral under an extended symmetry, in such a way that the VL fermions get their mass when this symmetry is broken, similar to how right-handed neutrinos get mass in Majoron models where lepton number is spontaneously broken. However, the implementation of such a mechanism in our theory lies beyond the scope of this work.} of new physics (NP)
above electroweak symmetry breaking, as shown in Fig.~\ref{fig:spectrum}.
Indeed, we can explain the whole flavour structure of the SM in terms
of just three small parameters, 
\begin{equation}
\frac{\langle\phi_{23}\rangle}{M_{23}}\sim\lambda^{3}\,,\qquad\frac{\langle\phi_{12}\rangle}{M_{12,13}}\sim\lambda\,,\qquad\frac{\langle\phi_{23}\rangle}{M_{12,13}}\sim\lambda^{5}\,.\label{eq:ThreeSmallParameters}
\end{equation}
The third ratio $\langle\phi_{23}\rangle/M_{12,13}\sim\lambda^{5}$,
which only appears in the (1,1) entry of the Yukawa matrices, provides
an extra (welcomed) suppression for first family fermion masses. Moreover,
this predicts a highly non-generic\footnote{Notice that this is a highly non-generic feature in this class of models.
Typically, alternative theories only contain ratios of the form $\langle\phi_{23}\rangle/M_{23}$
and $\langle\phi_{12}\rangle/M_{12}$ with no connection between the
23 and 12 VEVs.} relation between the two VEV scales as follows (up to $\mathcal{O}(1)$
variation), 
\begin{equation}
\langle\phi_{23}\rangle/\langle\phi_{12}\rangle=v_{23}/v_{12}\sim\lambda^{4}\,,\label{eq:VEV_relation}
\end{equation}
and also, together with the first ratio of Eq.~\eqref{eq:ThreeSmallParameters},
it suggests that $M_{23}\sim v_{12}$.
A typical benchmark compatible
with experimental searches then is, as depicted in Fig.~\ref{fig:spectrum},
\begin{equation}
v_{23}\sim\mathcal{O}(10\,\mathrm{TeV}),\quad v_{12}\sim M_{U_{23},D_{23},E_{23}}\sim\mathcal{O}(10^{3}\,\mathrm{TeV}),\quad M_{U_{12,13},D_{12,13},E_{12,13}}\sim\mathcal{O}(10^{4}\,\mathrm{TeV})\,,
\end{equation}
where, thanks to the non-generic relation of Eq.~(\ref{eq:VEV_relation}),
testing experimentally any of these scales effectively tests all of
them (and in particular, setting bounds on any of them also sets bounds
on the rest). Notice that our model then translates the complicated
flavour structure of the SM into just three simple and physical
scales of new physics above electroweak symmetry breaking, as shown in Fig.~\ref{fig:spectrum}.

However, we do not expect the masses of different messengers
nor the VEVs of different hyperons to be exactly degenerate - they may naturally
vary by $\mathcal{O}(1)$ factors. Therefore, we take advantage of
these natural $\mathcal{O}(1)$ variations with respect to Eq.~(\ref{eq:ThreeSmallParameters})
to explain the slightly more accentuated mass hierarchies of the up
sector and the tinier mass of the electron with respect to first family
quarks. More specifically, we assume the following numerical values
for the ratios appearing in the Yukawa matrices, which are consistent
with $\mathcal{O}(1)$ variations from the spectrum of Fig.~\ref{fig:spectrum}
(or similarly $\mathcal{O}(1)$ variations from the three fixed ratios
of Eq.~(\ref{eq:ThreeSmallParameters})), 
\begin{equation}
{\displaystyle \frac{\langle{\phi}_{\ell23}\rangle}{M_{U_{23}}}}\sim\lambda^{4}\,,\qquad{\displaystyle \frac{\langle{\phi}_{q23}\rangle}{M_{U_{23}}}}\sim\lambda^{3}\,,\qquad\frac{\langle{\phi}_{\ell12}\rangle}{M_{U_{12,13}}}\sim\frac{\langle{\phi}_{q12}\rangle}{M_{U_{12,13}}}\sim\lambda^{2}\,,
\end{equation}
\begin{equation}
{\displaystyle \frac{\langle{\phi}_{\ell23}\rangle}{M_{D_{23}}}}\sim\lambda^{3}\,,\qquad{\displaystyle \frac{\langle{\phi}_{q23}\rangle}{M_{D_{23}}}}\sim\lambda^{2}\,,\qquad\frac{\langle{\phi}_{\ell12}\rangle}{M_{D_{12,13}}}\sim\frac{\langle{\phi}_{q12}\rangle}{M_{D_{12,13}}}\sim\lambda\,,
\end{equation}
\begin{equation}
{\displaystyle \frac{\langle{\phi}_{\ell23}\rangle}{M_{E_{23}}}}\sim\lambda^{3}\,,\qquad\frac{\langle{\phi}_{\ell12}\rangle}{M_{E_{12,13}}}\sim\lambda^{2}\,,
\end{equation}
and the (1,1) entries of the effective Yukawa matrices are further
suppressed by fixed ratios that involve 23-VEVs in the numerator and
12-masses in the denominator, 
\begin{equation}
{\displaystyle \frac{\langle{\phi}_{\ell23}\rangle}{M_{U_{12,13}}}}\sim\lambda^{5}\,,\qquad{\displaystyle \frac{\langle{\phi}_{\ell23}\rangle}{M_{D_{12,13}}}}\sim\lambda^{4}\,,\qquad{\displaystyle \frac{\langle{\phi}_{\ell23}\rangle}{M_{E_{12,13}}}}\sim\lambda^{5}\,.
\end{equation}
We then predict the approximate numerical textures (up to $\mathcal{O}(1)$
coefficients) for the effective Yukawa couplings 
\begin{flalign}
{\cal L} & =\begin{pmatrix}q_{1} & q_{2} & q_{3}\end{pmatrix}\begin{pmatrix}{\displaystyle \lambda^{7}} & {\displaystyle \lambda^{6}} & \lambda^{5}\\
\lambda^{8} & \lambda^{4} & \lambda^{3}\\
0 & 0 & 1
\end{pmatrix}\begin{pmatrix}u_{1}^{c}\\
u_{2}^{c}\\
u_{3}^{c}
\end{pmatrix}v_{\mathrm{SM}}\\
 & +\begin{pmatrix}q_{1} & q_{2} & q_{3}\end{pmatrix}\begin{pmatrix}{\displaystyle \lambda^{5}} & {\displaystyle \lambda^{4}} & \lambda^{3}\\
\lambda^{5} & \lambda^{3} & \lambda^{2}\\
0 & 0 & 1
\end{pmatrix}\begin{pmatrix}d_{1}^{c}\\
d_{2}^{c}\\
d_{3}^{c}
\end{pmatrix}\lambda^{2}v_{\mathrm{SM}}\\
 & +\begin{pmatrix}\ell_{1} & \ell_{2} & \ell_{3}\end{pmatrix}\begin{pmatrix}\lambda^{7} & \lambda^{5} & \lambda^{5}\\
\lambda^{7} & {\displaystyle \lambda^{3}} & {\displaystyle \lambda^{3}}\\
0 & 0 & 1
\end{pmatrix}\begin{pmatrix}e_{1}^{c}\\
e_{2}^{c}\\
e_{3}^{c}
\end{pmatrix}\lambda^{2}v_{\mathrm{SM}}+\mathrm{h.c.}\,,
\end{flalign}
which can explain the flavour structure of the SM with all fundamental
couplings being $\mathcal{O}(1)$.

Notice that according to the above textures, CKM mixing in this model
originates mostly from the down sector. However, if one assumes for simplicity degenerate VEVs and masses as in Eq.~(\ref{eq:ThreeSmallParameters}), then the Yukawa textures would be similar (up to an overall normalisation factor) in both the up and down sectors - then the alignment of the CKM depends
on the off-diagonal $c_{ij}^{u,d}$ coefficients, as in the model of Ref.~\cite{FernandezNavarro:2023hrf}. In this case, all the coefficients $c_{ij}^{u,d,e}$ would still be of $\mathcal{O}(1)$, with the exception of $c_{11}^{u}$ and $c_{11}^{d}$ which would need to be of $\mathcal{O}(0.1)$ and of
$\mathcal{O}(4)$, respectively. This
is completely acceptable as $c_{11}^{u,d}$ are given as a product
of three fundamental couplings in the complete theory.

On the other hand, the prediction of significant 1-2 RH quark mixing,
especially in the down sector, is of great importance for the phenomenology
of the heaviest $Z'_{12}$ boson. Similarly, the presence of CKM-like
mixing in the charged lepton sector implies that the two $Z'$ bosons
of the model potentially mediate charged lepton flavour-violating processes (CLFV)
at acceptable rates, as discussed later in Section~\ref{sec:Phenomenology}. Most of the PMNS mixing is however generated from the neutrino sector, as discussed later in Section~\ref{sec:Neutrinos}.

\subsection{Model 2}
\label{sec:Model2}

Now we envisage an alternative complete and renormalisable model which contains less VL fermion
messengers by exchanging several of them by heavy Higgs doublets.
Here we exchange the VL fermions of $D$-type and $E$-type by the pair of Higgs doublets $H_{1}^{d}\sim(\mathbf{1,2})_{(-\frac{1}{2},0,0)}$
and $H_{2}^{d}\sim(\mathbf{1,2})_{(0,-\frac{1}{2},0)}$, as shown in Table~\ref{tab:UVmodel2}. These messengers
are motivated by the tri-hypercharge model of Ref.~\cite{FernandezNavarro:2023hrf}
that originates from a gauge unified framework. However, we still need to consider VL quarks to generate the
effective Yukawa operators that provide quark mixing. We could use
the VL quark doublets $Q_{i}\sim(\mathbf{3,2})_{\frac{1}{6}_{i}}$
as in \cite{FernandezNavarro:2023hrf}, with $i=1,2,3$. Nevertheless, here
we shall stick with the more minimal set of up-type VL quark singlets $U_{12}$, $U_{13}$ and $U_{23}$ that were already present in Model~1. On top of generating quark mixing, the $U_{ij}$ VL quarks will also generate the small masses of light up-quarks, and
will provide unique phenomenological predictions as we shall see in Section~\ref{sec:Phenomenology}.

\begin{table}
\begin{centering}
\begin{tabular}{ccccc}
\toprule 
Field  & $U(1)_{Y_{1}}$  & $U(1)_{Y_{2}}$  & $U(1)_{Y_{3}}$  & $SU(3)_{c}\times SU(2)_{L}$\tabularnewline
\midrule 
$\;\;\,H_{3}^{u,d}$  & 0  & 0  & $\pm1/2$  & $(\mathbf{1,2})$\tabularnewline
$H_{2}^{d}$  & 0  & $-1/2$  & 0  & $(\mathbf{1,2})$\tabularnewline
$H_{1}^{d}$  & $-1/2$  & 0  & 0  & $(\mathbf{1,2})$\tabularnewline
\midrule 
$\phi_{q_{12}}$  & -1/6  & 1/6  & 0  & $(\mathbf{1,1})$\tabularnewline
$\phi_{\ell_{12}}$  & 1/2  & -1/2  & 0  & $(\mathbf{1,1})$\tabularnewline
$\phi_{q_{23}}$  & 0  & -1/6  & 1/6  & $(\mathbf{1,1})$\tabularnewline
$\phi_{\ell_{23}}$  & 0  & 1/2  & -1/2  & $(\mathbf{1,1})$\tabularnewline
\midrule 
\rowcolor{yellow!10}$U_{12}$  & -1/6  & -1/2  & 0  & $(\mathbf{\overline{3},1})$\tabularnewline
\rowcolor{yellow!10}$U_{13}$  & -1/6  & 0  & -1/2  & $(\mathbf{\overline{3},1})$\tabularnewline
\rowcolor{yellow!10}$U_{23}$  & 0  & -1/6  & -1/2  & $(\mathbf{\overline{3},1})$\tabularnewline
\midrule
\rowcolor{yellow!10}$N_{12}$  & 1/2  & -1/2  & 0  & $(\mathbf{1,1})$\tabularnewline
\rowcolor{yellow!10}$N_{13}$  & 1/2  & 0  & -1/2  & $(\mathbf{1,1})$\tabularnewline
\rowcolor{yellow!10}$N_{23}$  & 0  & 1/2  & -1/2  & $(\mathbf{1,1})$\tabularnewline
\bottomrule
\end{tabular}
\par\end{centering}
\caption{Scalar and vector-like fermion content of
Model 2. As in Table~\ref{tab:UVmodel1}, for vector-like fermions (highlighted in yellow),
the conjugate partners are also included but not explicitly shown. \label{tab:UVmodel2}}
\end{table}

With respect to Model 1, Model 2 contains a smaller number of total degrees
of freedom and representations, but also more scalars fields and hence
more terms in the scalar potential. We refer to Appendix~\ref{app:ScalarPotential}, where we discuss the scalar potential of the two models and briefly comment as well on the stability of the spectrum in Fig.~\ref{fig:spectrum} under radiative corrections.

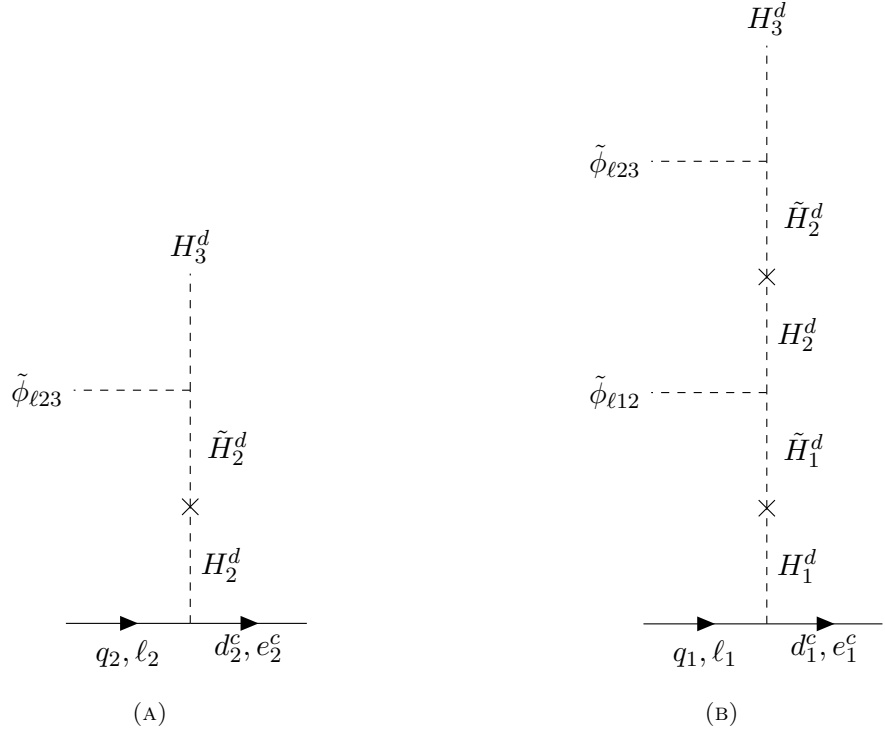
\begin{figure}[t]
\centering{}\subfloat[]{\resizebox{.26\textwidth}{!}{ \begin{tikzpicture}
	\begin{feynman}
		\vertex (a);
		\vertex [right=16mm of a] (b);
		\vertex [right=of b] (c);
		\vertex [above=of b] (f1);
		\vertex [above=of f1] (f2);
		\vertex [left=of f2] (g1) {\(\tilde{\phi}_{\ell 23}\)};
		\vertex [above=of f2] (g2) {\(H^{d}_{3}\)};;
		\diagram* {
			(a) -- [fermion, edge label'=\(q_{2}{,}\,\ell_{2}\), inner sep=6pt] (b) -- [fermion, edge label'=\(d^{c}_{2}{,}\,e^{c}_{2}\)] (c),
			(b) -- [scalar, edge label'=\(H_{2}^{d}\)] (f1),
			(f1) -- [scalar, edge label'=\(\tilde{H}_{2}^{d}\),inner sep=6pt, insertion=0] (f2),
            (f2) -- [scalar] (g1),
            (f2) -- [scalar] (g2),
	};
	\end{feynman}
\end{tikzpicture} }

}$\qquad\qquad\qquad\qquad$\subfloat[]{\begin{centering}
\resizebox{.25\textwidth}{!}{ \begin{tikzpicture}
	\begin{feynman}
		\vertex (a);
		\vertex [right=16mm of a] (b);
		\vertex [right=of b] (c);
		\vertex [above=of b] (f1);
		\vertex [above=of f1] (f2);
		\vertex [left=of f2] (g1) {\(\tilde{\phi}_{\ell 12}\)};
		\vertex [above=of f2] (g2);
		\vertex [above=of g2] (g3);
		\vertex [left=of g3] (i1) {\(\tilde{\phi}_{\ell 23}\)};
		\vertex [above=of g3] (i2) {\(H^{d}_{3}\)};
		\diagram* {
			(a) -- [fermion, edge label'=\(q_{1}{,}\,\ell_{1}\), inner sep=6pt] (b) -- [fermion, edge label'=\(d^{c}_{1}{,}\,e^{c}_{1}\)] (c),
			(b) -- [scalar, edge label'=\(H_{1}^{d}\)] (f1),
			(f1) -- [scalar, edge label'=\(\tilde{H}_{1}^{d}\),inner sep=6pt, insertion=0] (f2),
			(f2) -- [scalar] (g1),
			(f2) -- [scalar, edge label'=\(H_{2}^{d}\)] (g2),
			(g2) -- [scalar, edge label'=\(\tilde{H}_{2}^{d}\),inner sep=6pt, insertion=0] (g3),
            (g3) -- [scalar] (i1),
            (g3) -- [scalar] (i2),
	};
	\end{feynman}
\end{tikzpicture}} 
\par\end{centering}
}\caption[]{Diagrams at the origin of light down-quarks and charged
lepton masses in Model 2. The up-sector is described by diagrams such
as those of Fig.~\ref{fig:DiagramsModel1}. \label{fig:DiagramsModel2}}
\end{figure}

We remark that the full set of renormalisable Yukawa couplings in
the up sector is also described in this model by Eq.~(\ref{eq:up_FullCouplings}).
In contrast, the renormalisable Yukawa couplings in the down and charged
lepton sector are simply given by 
\begin{equation}
\mathcal{L}_{Y}^{d,e}=\sum_{i=1,2,3}y_{i}^{d}q_{i}H_{i}^{d}d_{i}^{c}+y_{i}^{e}\ell_{i}H_{i}^{d}e_{i}^{c}+\mathrm{h.c.}
\end{equation}
Notice however that the Higgs doublets $H_{1}^{d}$ and $H_{2}^{d}$
are assumed to \textit{not} get a VEV at the electroweak scale, but
instead we assume them to be very heavy and act as heavy messengers
of the diagonal entries in the effective Yukawa couplings of down
quarks and charged leptons, as shown in Fig.~\ref{fig:DiagramsModel2}.
For this we use the $d=3$ couplings $f_{12}^{dd}\widetilde{H}_{1}^{d}H_{2}^{d}\widetilde{\phi}_{\ell12}$
and $f_{23}^{dd}\widetilde{H}_{2}^{d}H_{3}^{d}\widetilde{\phi}_{\ell23}$
that appear in the scalar potential and couple the heavy Higgs doublets
to the hyperons, see Appendix~\ref{app:ScalarPotential}. 

After integrating out the heavy messengers, the effective Yukawa couplings
for chiral fermions take the form 
\begin{flalign}
{\cal L}_{d,e} & =\begin{pmatrix}q_{1} & q_{2} & q_{3}\end{pmatrix}\mathrm{diag}\left(c_{11}^{d}{\displaystyle \frac{\tilde{\phi}_{\ell12}}{M_{H_{1}^{d}}}}{\displaystyle \frac{\tilde{\phi}_{\ell23}}{M_{H_{2}^{d}}}},\,{\displaystyle c_{22}^{d}\frac{\tilde{\phi}_{\ell23}}{M_{H_{2}^{d}}}},\,c_{33}^{d}\right)\begin{pmatrix}d_{1}^{c}\\
d_{2}^{c}\\
d_{3}^{c}
\end{pmatrix}H_{3}^{d}\\
 & +\begin{pmatrix}\ell_{1} & \ell_{2} & \ell_{3}\end{pmatrix}\mathrm{diag}\left(c_{11}^{e}{\displaystyle \frac{\tilde{\phi}_{\ell12}}{M_{H_{1}^{d}}}}{\displaystyle \frac{\tilde{\phi}_{\ell23}}{M_{H_{2}^{d}}}},\,{\displaystyle c_{22}^{e}\frac{\tilde{\phi}_{\ell23}}{M_{H_{2}^{d}}}},\,c_{33}^{e}\right)\begin{pmatrix}e_{1}^{c}\\
e_{2}^{c}\\
e_{3}^{c}
\end{pmatrix}H_{3}^{d}\,+\mathrm{h.c.}\,,
\end{flalign}
where 
\begin{equation}
c^{d,e}=\mathrm{diag}\left(y_{1}^{d,e}\frac{f_{12}^{dd}}{M_{H_{1}^{d}}}\frac{f_{23}^{dd}}{M_{H_{2}^{d}}},\,y_{2}^{d,e}\frac{f_{23}^{dd}}{M_{H_{2}^{d}}},\,y_{3}^{d,e}\right)
\end{equation}
and we assume for simplicity $f_{12}^{dd}\sim M_{H_{1}^{d}}$ and
$f_{23}^{dd}\sim M_{H_{2}^{d}}$. In contrast,
the up-quark sector remains described by Eq.~(\ref{eq:UpSectorYukawa}).

We can see that in this model the down-quark and charged lepton Yukawa
matrices are diagonal at tree-level, therefore CKM and PMNS mixing fully originate
from the up and neutrino sectors, respectively. The up-quark
Yukawa couplings take a similar form as in Model 1, and we can similarly
describe all the flavour structure of the SM in terms of three naturally
small parameters as in Eq.~(\ref{eq:ThreeSmallParameters}), obtaining three simple NP scales above electroweak symmetry
breaking as shown in Fig.~\ref{fig:spectrum}. However, here we shall
take advantage of the $\mathcal{O}(1)$ variation of the VL fermion
masses and VEVs of the hyperons (as discussed in Section~\ref{sec:Model1})
to choose slightly different ratios as 
\begin{equation}
{\displaystyle \frac{\langle{\phi}_{\ell23}\rangle}{M_{U_{23}}}}\sim\lambda^{3}\,,\qquad{\displaystyle \frac{\langle{\phi}_{q23}\rangle}{M_{U_{23}}}}\sim\lambda^{2}\,,\qquad\frac{\langle{\phi}_{\ell12}\rangle}{M_{U_{12,13}}}\sim\lambda^{2}\,,\qquad\frac{\langle{\phi}_{q12}\rangle}{M_{U_{12,13}}}\sim\lambda\,,
\end{equation}
and the (1,1) entry of the effective up-Yukawa matrix is further suppressed
by the fixed ratio,
\begin{equation}
{\displaystyle \frac{\langle{\phi}_{\ell23}\rangle}{M_{U_{12,13}}}}\sim\lambda^{5}\,.
\end{equation}
This choice is consistent with the spectrum of Fig.~\ref{fig:spectrum}
up to $\mathcal{O}(1)$ variations, now with the addition
of the two heavy Higgs doublets that live with the 23-messengers and
with the 12,13-messengers, respectively. This provides the
following values for the ratios involving the heavy Higgs
doublets, 
\begin{equation}
{\displaystyle \frac{\langle{\phi}_{\ell23}\rangle}{M_{H_{2}^{d}}}}\sim\lambda^{3}\,,\qquad\frac{\langle{\phi}_{\ell12}\rangle}{M_{H_{1}^{d}}}\sim\lambda^{3}\,.
\end{equation}
Notice that, as in Model 1, the presence of the ratio $\phi_{\ell23}/M_{U_{12}}$
provides a highly non-generic and predictive connection between both
VEVs as per Eq.~(\ref{eq:VEV_relation}), namely 
\begin{equation}
\langle\phi_{\ell23}\rangle/\langle\phi_{\ell12}\rangle=v_{23}/v_{12}\sim\lambda^{3}\sim0.01\,.
\end{equation}
Therefore, a typical benchmark for the scales of Model 2 compatible with phenomenology is 
\begin{equation}
v_{23}\sim\mathcal{O}(10\,\mathrm{TeV}),\quad v_{12}\sim M_{U_{23},H_{2}^{d}}\sim\mathcal{O}(10^{3}\,\mathrm{TeV})\,,\quad M_{U_{12,13},H_{1}^{d}}\sim\mathcal{O}(10^{5}\,\mathrm{TeV})\,.
\end{equation}
We conclude this section with the approximate textures (up to $\mathcal{O}(1)$
coefficients) predicted for the effective Yukawa couplings by Model 2,
\begin{flalign}
{\cal L} & =\begin{pmatrix}q_{1} & q_{2} & q_{3}\end{pmatrix}\begin{pmatrix}{\displaystyle \lambda^{7}} & {\displaystyle \lambda^{4}} & \lambda^{3}\\
\lambda^{6} & \lambda^{3} & \lambda^{2}\\
0 & 0 & 1
\end{pmatrix}\begin{pmatrix}u_{1}^{c}\\
u_{2}^{c}\\
u_{3}^{c}
\end{pmatrix}v_{\mathrm{SM}}\\
 & +\begin{pmatrix}q_{1} & q_{2} & q_{3}\end{pmatrix}\begin{pmatrix}\lambda^{6} & 0 & 0\\
0 & \lambda^{3} & 0\\
0 & 0 & 1
\end{pmatrix}\begin{pmatrix}d_{1}^{c}\\
d_{2}^{c}\\
d_{3}^{c}
\end{pmatrix}\lambda^{2}v_{\mathrm{SM}}\\
 & +\begin{pmatrix}\ell_{1} & \ell_{2} & \ell_{3}\end{pmatrix}\begin{pmatrix}\lambda^{6} & 0 & 0\\
0 & {\displaystyle \lambda^{3}} & {\displaystyle 0}\\
0 & 0 & 1
\end{pmatrix}\begin{pmatrix}e_{1}^{c}\\
e_{2}^{c}\\
e_{3}^{c}
\end{pmatrix}\lambda^{2}v_{\mathrm{SM}}+\mathrm{h.c.}
\end{flalign}
Since here CKM mixing originates from the
up sector and the charged lepton matrix is diagonal at tree-level,
the leading flavour phenomenology involves FCNCs in the up sector,
unlike Model 1 where the leading phenomenology
involves FCNCs in the down sector along with CLFV
processes.

\section{Neutrinos}
\label{sec:Neutrinos}

As first discussed in \cite{FernandezNavarro:2023rhv}, the tri-hypercharge symmetry imposes selection
rules on the Weinberg operator that potentially complicate the explanation
of large neutrino mixing. Nevertheless, this problem can be solved
via minimal extensions of the traditional seesaw mechanism. With respect
to \cite{FernandezNavarro:2023rhv}, here we propose a different, more minimal model for the neutrino
sector which does not require the addition of any extra scalar beyond
those already present in Models 1 and 2.

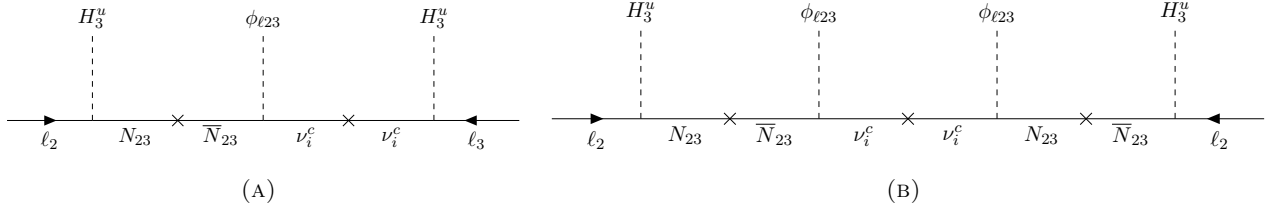
\begin{figure}[t]
\subfloat[]{\resizebox{.42\textwidth}{!}{ \begin{tikzpicture}
	\begin{feynman}
		\vertex (a);
		\vertex [right=16mm of a] (b);
		\vertex [right=16mm of b] (c);
		\vertex [right=16mm of c] (d);
		\vertex [right=16mm of d] (e);
		\vertex [right=16mm of e] (f);
		\vertex [right=16mm of f] (g);
		\vertex [above=16mm of b] (f1) {\(H^{u}_{3}\)};
		\vertex [above=16mm of d] (f2) {\(\phi_{\ell 23}\)};
		\vertex [above=16mm of f] (f3) {\(H^{u}_{3}\)};
		\diagram* {
			(a) -- [fermion, edge label'=\(\ell_{2}\), inner sep=6pt] (b) -- [scalar] (f1),
			(b) -- [edge label'=\(N_{23}\)] (c),
			(c) -- [edge label'=\(\overline{N}_{23}\),inner sep=2pt, insertion=0] (d) -- [scalar] (f2),
			(d) -- [edge label'=\(\nu^{c}_{i}\)] (e),
			(e) -- [edge label'=\(\nu^{c}_{i}\),inner sep=3.5pt, insertion=0] (f) -- [scalar] (f3),
			(f) -- [anti fermion, edge label'=\(\ell_{3}\), inner sep=6pt] (g),
	};
	\end{feynman}
\end{tikzpicture} }

}$\;$\subfloat[]{\resizebox{.58\textwidth}{!}{ \begin{tikzpicture}
	\begin{feynman}
		\vertex (a);
		\vertex [right=16mm of a] (b);
		\vertex [right=16mm of b] (c);
		\vertex [right=16mm of c] (d);
		\vertex [right=16mm of d] (e);
		\vertex [right=16mm of e] (f);
		\vertex [right=16mm of f] (g);
		\vertex [right=16mm of g] (h);
		\vertex [right=16mm of h] (i);
		\vertex [above=16mm of b] (f1) {\(H^{u}_{3}\)};
		\vertex [above=16mm of d] (f2) {\(\phi_{\ell 23}\)};
		\vertex [above=16mm of f] (f3) {\(\phi_{\ell 23}\)};
		\vertex [above=16mm of h] (f4) {\(H^{u}_{3}\)};
		\diagram* {
			(a) -- [fermion, edge label'=\(\ell_{2}\), inner sep=6pt] (b) -- [scalar] (f1),
			(b) -- [edge label'=\(N_{23}\)] (c),
			(c) -- [edge label'=\(\overline{N}_{23}\),inner sep=2pt, insertion=0] (d) -- [scalar] (f2),
			(d) -- [edge label'=\(\nu^{c}_{i}\)] (e),
			(e) -- [edge label'=\(\nu^{c}_{i}\),inner sep=3pt, insertion=0] (f) -- [scalar] (f3),
			(f) -- [edge label'=\(N_{23}\)] (g),
			(g) -- [edge label'=\(\overline{N}_{23}\),inner sep=2pt, insertion=0] (h) -- [scalar] (f4),
			(h) -- [anti fermion, edge label'=\(\ell_{2}\), inner sep=6pt] (i),
	};
	\end{feynman}
\end{tikzpicture} }

}

\caption[]{Illustrative diagrams for the generation of effective Majorana masses for active neutrinos, where $i=1,2$.
\label{fig:Neutrino_diagrams}}
\end{figure}

In order to describe neutrino masses and mixing, we use the two complete
singlet neutrinos $\nu_{1,2}^{c}$ and the two hyperons $\phi_{\ell23}$
and $\phi_{\ell12}$ already included in Table~\ref{tab:Field_content}, along with thee vector-like neutrinos
$N_{12}$, $N_{13}$ and $N_{23}$ common to both Models 1 and 2 as shown in Tables~\ref{tab:UVmodel1} and \ref{tab:UVmodel2}. The rationale is that the vector-like neutrinos,
thanks to the presence of the hyperons $\phi_{\ell23}$
and $\phi_{\ell12}$, can mediate couplings
between the light lepton doublets and the (very heavy) singlet neutrinos
$\nu_{1,2}^{c}$, as shown in Fig.~\ref{fig:Neutrino_diagrams}, which would otherwise be forbidden by the selection rules.

However, the diagrammatic approach of Fig.~\ref{fig:Neutrino_diagrams} requires to employ the
mass insertion approximation where $M_{N_{23}}\gg\langle\phi_{\ell23}\rangle$,
hence typically predicting suppressed neutrino mixing in conflict
with measured values \cite{deSalas:2020pgw,Gonzalez-Garcia:2021dve}. Nevertheless, the diagrams in Fig.~\ref{fig:Neutrino_diagrams} give
the intuition that if $M_{N_{23}}\sim\langle\phi_{\ell23}\rangle$
(and similarly $M_{N_{12,13}}\sim\langle\phi_{\ell12}\rangle$), then
large neutrino mixing can be obtained. In order to verify this, we need
to disregard the mass insertion approximation, because in the regime
$M_{N_{23}}\sim\langle\phi_{23}\rangle$ the EFT suggested by the diagrams in
Fig.~\ref{fig:Neutrino_diagrams} is not consistent.

The way to go is to introduce formally the three messenger neutrinos
$N_{12}$, $N_{13}$ and $N_{23}$ in the theory and employ the seesaw
formula. We can construct the Dirac and Majorana mass matrices in
the usual way by writing all gauge invariant couplings as 
\begin{equation}
m_{D}=\left(\begin{array}{c|cccccccc}
 & \overline{N}_{12} & \overline{N}_{13} & \overline{N}_{23} & N_{12} & N_{13} & N_{23} & \nu_{1}^{c} & \nu_{2}^{c}\\
\hline \ell_{1} & 0 & 0 & 0 & 0 & y_{13}H_{3}^{u} & 0 & 0 & 0\\
\ell_{2} & 0 & 0 & 0 & 0 & 0 & y_{24}H_{3}^{u} & 0 & 0\\
\ell_{3} & 0 & 0 & 0 & 0 & 0 & 0 & y_{35}H_{3}^{u} & y_{36}H_{3}^{u}
\end{array}\right)\,,
\end{equation}
\begin{equation}
M_{N}=\left(\begin{array}{c|cccccccc}
 & \overline{N}_{12} & \overline{N}_{13} & \overline{N}_{23} & N_{12} & N_{13} & N_{23} & \nu_{1}^{c} & \nu_{2}^{c}\\
\hline \overline{N}_{12} & 0 & 0 & 0 & m_{N_{12}} & x_{15}\widetilde{\phi}_{\ell23} & 0 & z_{17}\phi_{\ell12} & z_{18}\phi_{\ell12}\\
\overline{N}_{13} & 0 & 0 & 0 & x_{24}\phi_{\ell23} & m_{N_{13}} & x_{26} \phi_{\ell12} & 0 & 0\\
\overline{N}_{23} & 0 & 0 & 0 & 0 & x_{35} \widetilde{\phi}_{\ell12} & m_{N_{23}} & z_{37}\phi_{\ell23} & z_{38}\phi_{\ell23}\\
N_{12} & m_{N_{12}} & x_{24}\phi_{\ell23} & 0 & 0 & 0 & 0 & z_{47}\widetilde{\phi}_{\ell12} & z_{48}\widetilde{\phi}_{\ell12}\\
N_{13} & x_{15}\widetilde{\phi}_{\ell23} & m_{N_{13}} & x_{35}\widetilde{\phi}_{\ell12} & 0 & 0 & 0 & 0 & 0\\
N_{23} & 0 & x_{26}\phi_{\ell12} & m_{N_{23}} & 0 & 0 & 0 & z_{67}\widetilde{\phi}_{\ell23} & z_{68}\widetilde{\phi}_{\ell23}\\
\nu_{1}^{c} & z_{17}\phi_{\ell12} & 0 & z_{37}\phi_{\ell23} & z_{47}\widetilde{\phi}_{\ell12} & 0 & z_{67}\widetilde{\phi}_{\ell23} & M_{11} & M_{12}\\
\nu_{2}^{c} & z_{18}\phi_{\ell12} & 0 & z_{38}\phi_{\ell23} & z_{48}\widetilde{\phi}_{\ell12} & 0 & z_{68}\widetilde{\phi}_{\ell23} & M_{12} & M_{22}
\end{array}\right)\,,
\end{equation}
where we can take advantage of a global $U(2)$ symmetry relating $\nu_{1}^{c}$
and $\nu_{2}^{c}$ to set $M_{12}=0$ without loss of generality.
The full neutrino mass matrix is given as
\begin{equation}
M_{\nu}=\left(\begin{array}{cc}
0 & m_{D}\\
m_{D}^{\mathrm{T}} & M_{N}
\end{array}\right)\,.\label{eq:Full_Mnu}
\end{equation}
Since $m_{D}$ only contains couplings to the Higgs doublet $H^{u}_{3}$ that breaks electroweak symmetry, while $M_{N}$ in all cases contains heavy scales in the multi-TeV region or above, we are in the regime $m_{D}\ll M_{N}$ where we can safely apply the seesaw formula.

But first we note that after exchanging the hyperons by their VEVs $\langle\phi_{\ell12}\rangle\sim v_{12}$
and $\langle\phi_{\ell23}\rangle\sim v_{23}$, the submatrix in $M_N$ that corresponds to the VL messengers contains two eigenvalues at $v_{12}$
and one at $v_{23}$ (as we expect from the spectrum in Fig.~\ref{fig:spectrum})
only for the choice of bare masses $m_{N_{12}}\sim v_{23}$, $m_{N_{13}}\sim v_{23}$
and $m_{N_{23}}\sim v_{12}$, namely
\begin{equation}
\left(\begin{array}{c|ccc}
 & N_{12} & N_{13} & N_{23}\\
\hline \overline{N}_{12} & v_{23} & x_{15}v_{23} & 0\\
\overline{N}_{13} & x_{24}v_{23} & v_{23} & x_{26}v_{12}\\
\overline{N}_{23} & 0 & x_{35}v_{12} & v_{12}
\end{array}\right)\overset{\mathrm{diag}}{\longrightarrow}\left(\begin{array}{c|ccc}
 & \hat{N}_{23} & \hat{N}_{13} & \hat{N}_{12}\\
\hline \hat{\overline{N}}_{23} & v_{23} & 0 & 0\\
\hat{\overline{N}}_{13} & 0 & v_{12} & 0\\
\hat{\overline{N}}_{12} & 0 & 0 & v_{12}
\end{array}\right)\,,
\end{equation}
where we have assumed $\mathcal{O}(1)$ coefficients $x_{ij}$ and
$v_{23}\ll v_{12}$ as predicted from Eq.~\eqref{eq:VEV_relation}. The choice of bare masses
$m_{N_{ij}}$ above may seem contrary to the prescription $M_{N_{23}}\sim v_{23}$
(and similarly $M_{N_{12,13}}\sim v_{12}$) that we anticipated from
the diagrammatic approach, but we need to note that $m_{N_{ij}}$
are not necessarily the physical masses
of the messenger neutrinos. Indeed, in the regime where the messenger
neutrinos and the VEVs of the hyperons are of the same order, the
mixing terms among $N_{ij}$ are very relevant. After taking them
into account, our choice of bare masses delivers $M_{N_{23}}\sim v_{23}$
and $M_{N_{12,13}}\sim v_{12}$, as the physical masses of the messenger
neutrinos. After applying the seesaw formula to extract the light neutrino
masses, we verify that messenger neutrinos at these scales mediate $\mathcal{O}(1)$
couplings between the light lepton doublets and the singlet neutrinos.
Indeed, we obtain that to leading order all the $v_{12}$ and $v_{23}$
scales cancel between the numerator and the denominator. Namely, we find
\begin{flalign}
 & m_{\nu}\simeq m_{D}M_{N}^{-1}m_{D}^{\mathrm{T}} \nonumber\\
 & =\left(\begin{array}{ccc}
{\displaystyle y_{13}^{2}x_{24}^{2}\frac{M_{11}z_{18}^{2}+M_{22}z_{17}^{2}}{M_{11}M_{22}x_{26}^{2}x_{35}^{2}}} & {\displaystyle -y_{13}y_{24}x_{24}^{2}\frac{M_{11}z_{18}^{2}+M_{22}z_{17}^{2}}{M_{11}M_{22}x_{26}^{2}x_{35}}} & {\displaystyle -y_{13}\frac{M_{11}z_{18}y_{36}+M_{22}z_{17}y_{35}}{M_{11}M_{22}x_{26}x_{35}}}\\
{\displaystyle -y_{13}y_{24}x_{24}^{2}\frac{M_{11}z_{18}^{2}+M_{22}z_{17}^{2}}{M_{11}M_{22}x_{26}^{2}x_{35}}} & {\displaystyle y_{24}^{2}x_{24}^{2}\frac{M_{11}z_{18}^{2}+M_{22}z_{17}^{2}}{M_{11}M_{22}x_{26}^{2}}} & {\displaystyle y_{24}x_{24}\frac{M_{11}z_{18}y_{36}+M_{22}z_{17}y_{35}}{M_{11}M_{22}x_{26}}}\\
{\displaystyle -y_{13}\frac{M_{11}z_{18}y_{36}+M_{22}z_{17}y_{35}}{M_{11}M_{22}x_{26}x_{35}}} & {\displaystyle y_{24}x_{24}\frac{M_{11}z_{18}y_{36}+M_{22}z_{17}y_{35}}{M_{11}M_{22}x_{26}}} & {\displaystyle \frac{M_{11}y_{36}^{2}+M_{22}y_{35}^{2}}{M_{11}M_{22}}}
\end{array}\right)v_{\mathrm{SM}}^{2}\nonumber \\
 & +\mathcal{O}\left(v_{\mathrm{SM}}^{2}\frac{M_{11,22}}{M_{11}M_{22}}\frac{v_{23}}{v_{12}}\right) \, .
\end{flalign}
If we assume $M_{11}\ll M_{22}$ as in sequential dominance \cite{King:1998jw,King:1999mb,King:2002nf}, and
$x_{ij}=z_{ij}=1$ for simplicity, then we have 
\begin{equation}
m_{\nu}\simeq m_{D}M_{N}^{-1}m_{D}^{\mathrm{T}}\approx\left(\begin{array}{ccc}
y_{13}^{2} & -y_{13}y_{24} & -y_{13}y_{35}\\
-y_{13}y_{24} & y_{24}^{2} & y_{24}y_{35}\\
-y_{13}y_{35} & y_{24}y_{35} & y_{35}^{2}
\end{array}\right)\frac{v_{\mathrm{SM}}^{2}}{M_{11}}
\end{equation}
This result is similar to what one would obtain in a minimal
type-I seesaw extension of the SM. Indeed the texture above predicts
two very light active neutrinos (and one massless) if $M_{11}\approx10^{15}$~GeV, and is capable to explain large neutrino mixing by fitting the $\mathcal{O}(1)$
coefficients to existing data. Notably, this implementation of the
seesaw mechanism requires no addition of extra scalars beyond those
already introduced to explain the flavour structure of the charged
fermion sector. Moreover, there is no need to impose small couplings
nor assuming that the scales $v_{12}$ or $v_{23}$ are very heavy
in order to achieve small neutrino masses (meaning that such scales
can be in the multi-TeV region), as is commonly assumed in the literature.

We remark that the three neutrino messengers $N_{12}$,
$N_{13}$ and $N_{23}$ obtain masses at the same scale of the $Z'_{12}$ and $Z'_{23}$ bosons to which they couple and which could live in the multi-TeV region, meaning that they could carry
meaningful phenomenology if such $Z'$ bosons are discovered.

\section{Phenomenology}
\label{sec:Phenomenology}

\subsection{Gauge boson couplings and complete fermion mixing} \label{sec:GaugeCouplings}

The phenomenology of the tri-hypercharge framework is mostly driven
by the $Z'_{12}$ and $Z'_{23}$ gauge bosons that arise from the
12 and 23 breaking, respectively, and to some extent also to the $Z$
boson, as firstly discussed in \cite{FernandezNavarro:2023rhv} where
the couplings of such gauge bosons to chiral fermions can be found.

The only difference with respect to the study in \cite{FernandezNavarro:2023rhv}
is the inclusion of the heavy messengers that mediate the effective
Yukawa couplings in the complete models presented here. Beyond dictating
the origin and size of chiral fermion mixing, which controls the size
and nature of potential flavour-changing neutral currents (FCNCs)
mediated by the gauge bosons, the vector-like fermion messengers also
couple to the gauge bosons and mix with chiral fermions, hence potentially
providing new sources of flavour-violation beyond those that originate
from chiral fermion mixing. The fermion couplings of the gauge bosons
may be represented as $6\times6$ matrices in flavour space, with
the vector-like fermion families $F_{12}$, $F_{13}$ and $F_{23}$
(with $F=U,D,E,N$ for Model 1, and $F=U,N$ only in Model 2) taken
as fourth, fifth and sixth family respectively. For simplicity, we
start by treating separately the couplings of chiral and vector-like
fermions with the gauge bosons, in order to write in the interaction
basis\footnote{Here we depart from our purely left-handed and two-component convention
to work in a four-component, left-right notation more familiar for
phenomenological studies. Note that the sign of the hypercharges of $f_{R}$ and $F_{L,R}$ fermions is opposite with respect to $f^{c}$ and VL fermions in Tables \ref{tab:UVmodel1} and \ref{tab:UVmodel2}.}
\begin{flalign}
\mathcal{L}_{\mathrm{gauge}} & \supset(\kappa_{ij}^{f_{L,R}}\overline{f}_{L,R}^{i}\gamma_{\mu}f_{L,R}^{j}+\kappa_{ab}^{F}\overline{F}_{L,R}^{a}\gamma_{\mu}F_{L,R}^{b})Z'^{\mu}_{12}+(\xi_{ij}^{f_{L,R}}\overline{f}_{L,R}^{i}\gamma_{\mu}f_{L,R}^{j}+\xi_{ab}^{F}\overline{F}_{L,R}^{a}\gamma_{\mu}F_{L,R}^{b})Z'^{\mu}_{23}\\
 & +(\omega_{ij}^{f_{L,R}}\overline{f}_{L,R}^{i}\gamma_{\mu}f_{L,R}^{j}+\omega_{ab}^{F}\overline{F}_{L,R}^{a}\gamma_{\mu}F_{L,R}^{b})Z^{\mu}\,,\nonumber 
\end{flalign}
where $f=u,d,e,\nu$ with $i,j=1,2,3$ and $a,b=4,5,6$. The coupling
matrices above are obtained from the covariant derivatives in \cite{FernandezNavarro:2023rhv}
\begin{flalign}
 & \kappa_{ij}^{f_{L,R}}=\mathrm{diag}(-Y_{1}^{f_{L,R}^{1}}g_{1}\sin\theta_{12}^{W},\;Y_{2}^{f_{L,R}^{2}}g_{2}\cos\theta_{12}^{W},\;0)\,,\\
 & \kappa_{ab}^{F}=\mathrm{diag}(-Y_{1}^{F_{12}}g_{1}\sin\theta_{12}^{W}+Y_{2}^{F_{12}}g_{2}\cos\theta_{12}^{W},\;-Y_{1}^{F_{13}}g_{1}\sin\theta_{12}^{W},\;Y_{2}^{F_{23}}g_{2}\cos\theta_{12}^{W})\,,\nonumber \\
 & \xi_{ij}^{f_{L,R}}=\mathrm{diag}(-Y_{12}^{f_{L,R}^{1}}g_{12}\sin\theta_{23}^{W},\;-Y_{12}^{f_{L,R}^{2}}g_{12}\sin\theta_{23}^{W},\;Y_{3}^{f_{L,R}^{3}}g_{3}\cos\theta_{23}^{W})\,,\nonumber \\
 & \xi_{ab}^{F}=\mathrm{diag}(-Y_{12}^{F_{12}}g_{12}\sin\theta_{23}^{W},\;-Y_{12}^{F_{13}}g_{12}\sin\theta_{23}^{W}+Y_{3}^{F_{13}}g_{3}\cos\theta_{23}^{W},\;-Y_{12}^{F_{23}}g_{12}\sin\theta_{23}^{W}+Y_{3}^{F_{23}}g_{3}\cos\theta_{23}^{W})\,,\nonumber \\
 & \omega_{ij}^{f_{L,R}}=(T_{f_{L,R}}^{3}g_{L}\cos\theta^{W}-Y^{f_{L,R}}g_{Y}\sin\theta^{W})\mathbb{I}_{\mathrm{3\times3}}\,,\qquad\omega_{ab}^{F}=-Y^{F}\,g_{Y}\sin\theta^{W}\mathbb{I}_{\mathrm{3\times3}}\,,\nonumber 
\end{flalign}
\begin{table}
\begin{centering}
\resizebox{\textwidth}{!}{
\begin{tabular}{llll}
\toprule 
$d_{L}$ sector & $d_{R}$ sector & $e_{L}$ sector & $e_{R}$ sector\tabularnewline
\midrule
\midrule 
$s_{12}^{d_{L}}\simeq\frac{c_{12}^{d}}{c_{22}^{d}}\frac{\langle\phi_{q12}\rangle}{M_{D_{13}}}\sim\lambda$ & $s_{12}^{d_{R}}\simeq\frac{c_{21}^{d}}{c_{22}^{d}}\frac{\langle\phi_{\ell12}\rangle}{M_{D_{12}}}\frac{\langle\phi_{q12}\rangle}{M_{D_{13}}}\sim\lambda^{2}$ & $s_{12}^{e_{L}}\simeq\frac{c_{12}^{e}}{c_{22}^{e}}\frac{\langle\phi_{\ell12}\rangle}{M_{E_{13}}}\sim\lambda^{2}$ & $s_{12}^{e_{R}}\simeq\frac{c_{21}^{e}}{c_{22}^{e}}\frac{\langle\phi_{\ell12}\rangle}{M_{E_{12}}}\frac{\langle\phi_{\ell12}\rangle}{M_{E_{13}}}\sim\lambda^{4}$\tabularnewline
\midrule 
$s_{13}^{d_{L}}\simeq\frac{c_{13}^{d}}{c_{33}^{d}}\frac{\langle\phi_{q12}\rangle}{M_{D_{13}}}\frac{\langle\phi_{q23}\rangle}{M_{D_{23}}}\sim\lambda^{3}$ & $s_{26}^{d_{R}}\simeq y_{62}^{d}\frac{\langle\phi_{\ell23}\rangle}{M_{D_{23}}}\sim\lambda^{3}$ & $s_{13}^{e_{L}}\simeq\frac{c_{13}^{e}}{c_{33}^{e}}\frac{\langle\phi_{\ell12}\rangle}{M_{E_{13}}}\frac{\langle\phi_{\ell23}\rangle}{M_{E_{23}}}\sim\lambda^{5}$ & $s_{26}^{e_{R}}\simeq y_{62}^{e}\frac{\langle\phi_{\ell23}\rangle}{M_{E_{23}}}\sim\lambda^{3}$\tabularnewline
\midrule 
$s_{23}^{d_{L}}\simeq\frac{c_{23}^{d}}{c_{33}^{d}}\frac{\langle\phi_{q23}\rangle}{M_{D_{23}}}\sim\lambda^{2}$ & $s_{36}^{d_{R}}\simeq y_{63}^{d}\frac{\langle\phi_{q23}\rangle}{M_{D_{23}}}\sim\lambda^{2}$ & $s_{23}^{e_{L}}\simeq\frac{c_{23}^{e}}{c_{33}^{e}}\frac{\langle\phi_{\ell23}\rangle}{M_{E_{23}}}\sim\lambda^{3}$ & $s_{36}^{e_{R}}\simeq y_{63}^{d}\frac{\langle\phi_{\ell23}\rangle}{M_{E_{23}}}\sim\lambda^{3}$\tabularnewline
\midrule 
$s_{26}^{d_{L}}\simeq y_{26}^{d}\frac{\langle H_{3}^{d}\rangle}{M_{D_{13}}}\sim10^{-6}$ & $s_{14}^{d_{R}}\simeq y_{41}^{d}\frac{\langle\phi_{\ell12}\rangle}{M_{D_{12}}}\sim\lambda$ & $s_{26}^{e_{L}}\simeq y_{26}^{e}\frac{\langle H_{3}^{d}\rangle}{M_{E_{13}}}\sim10^{-6}$ & $s_{14}^{e_{R}}\simeq y_{41}^{e}\frac{\langle\phi_{\ell12}\rangle}{M_{E_{12}}}\sim\lambda^{2}$\tabularnewline
\midrule 
$s_{15}^{d_{L}}\simeq y_{15}^{d}\frac{\langle H_{3}^{d}\rangle}{M_{D_{13}}}\sim10^{-7}$ & $s_{24}^{d_{R}}\simeq y_{42}^{d}\frac{\langle\phi_{q12}\rangle}{M_{D_{12}}}\sim\lambda$ & $s_{15}^{e_{L}}\simeq y_{15}^{e}\frac{\langle H_{3}^{d}\rangle}{M_{E_{13}}}\sim10^{-7}$ & $s_{24}^{e_{R}}\simeq y_{42}^{e}\frac{\langle\phi_{\ell12}\rangle}{M_{E_{12}}}\sim\lambda^{2}$\tabularnewline
\midrule 
$s_{56}^{d_{L}}\simeq y_{56}^{d}\frac{\langle\phi_{q12}\rangle}{M_{D_{13}}}\sim\lambda$ & $s_{56}^{d_{R}}\simeq y_{65}^{d}\frac{\langle\phi_{q12}\rangle}{M_{D_{13}}}\sim\lambda$ & $s_{56}^{e_{L}}\simeq y_{56}^{e}\frac{\langle\phi_{\ell12}\rangle}{M_{E_{13}}}\sim\lambda^{2}$ & $s_{56}^{e_{R}}\simeq y_{65}^{e}\frac{\langle\phi_{\ell12}\rangle}{M_{E_{13}}}\sim\lambda$\tabularnewline
\midrule 
$s_{45}^{d_{L}}\simeq y_{45}^{d}\frac{\langle\phi_{\ell23}\rangle}{M_{D_{12,13}}}\sim\lambda^{4}$ & $s_{45}^{d_{R}}\simeq y_{54}^{d}\frac{\langle\phi_{\ell23}\rangle}{M_{D_{12,13}}}\sim\lambda^{4}$ & $s_{45}^{e_{L}}\simeq y_{45}^{e}\frac{\langle\phi_{\ell23}\rangle}{M_{E_{12,13}}}\sim\lambda^{5}$ & $s_{45}^{e_{R}}\simeq y_{54}^{e}\frac{\langle\phi_{\ell23}\rangle}{M_{E_{12,13}}}\sim\lambda^{5}$\tabularnewline
\bottomrule
\end{tabular}
}
\par\end{centering}
\caption{Approximate charged fermion mixing in Model 1. The angles not shown are zero
at tree-level. We also neglect up-quark mixing, assuming that the
CKM mostly originates from the down sector as the textures in Eqs.~\eqref{eq:eff_Yukawa_up} and \eqref{eq:eff_Yukawa_down} suggest.\label{tab:MixingModel1}}
\end{table}
where $Y_{i}^{f_{L,R}^{i}}$ refers to the $i$-th hypercharge of
the $f_{L,R}^{i}$ fermion (and similarly for the vector-like fermions
$F$), $g_{Y}$ and $g_{L}$ are the conventional gauge couplings
of SM hypercharge and $SU(2)_{L}$ respectively, $T_{f_{L,R}}^{3}$
is the third component of weak isospin, $Y=Y_{1}+Y_{2}+Y_{3}$ is SM hypercharge, $\theta^{W}$ is the conventional
weak mixing angle and the weak-like mixing angles $\theta_{12,23}^{W}$
are defined as
\begin{equation}
\mathrm{sin}\theta_{12}^{W}=\frac{g_{1}}{\sqrt{g_{1}^{2}+g_{2}^{2}}},\qquad\qquad\qquad\mathrm{sin}\theta_{23}^{W}=\frac{g_{12}}{\sqrt{g_{12}^{2}+g_{3}^{2}}}\,,
\end{equation}
where the gauge couplings are related by the following matching conditions
\begin{equation}
g_{12}=\frac{g_{1}g_{2}}{\sqrt{g_{1}^{2}+g_{2}^{2}}},\qquad\qquad\qquad g_{Y}=\frac{g_{12}g_{3}}{\sqrt{g_{12}^{2}+g_{3}^{2}}}\,.
\end{equation}
We may use the matching condition with SM hypercharge to exchange
$g_{12}$ in favour of $g_{3}$ and $g_{Y}$ for the couplings of
$Z'_{23}$,
\begin{flalign*}
 & \xi_{ij}^{f_{L,R}}=\mathrm{diag}\left(-Y_{12}^{f_{L,R}^{1}}\frac{g_{Y}^{2}}{\sqrt{g_{3}^{2}-g_{Y}^{2}}},\;-Y_{12}^{f_{L,R}^{2}}\frac{g_{Y}^{2}}{\sqrt{g_{3}^{2}-g_{Y}^{2}}},\;Y_{3}^{f_{L,R}^{3}}\sqrt{g_{3}^{2}-g_{Y}^{2}}\right)\,,\\
 & \xi_{ab}^{F}=\mathrm{diag}\left(-Y_{12}^{F_{12}}\frac{g_{Y}^{2}}{\sqrt{g_{3}^{2}-g_{Y}^{2}}},\;-Y_{12}^{F_{13}}\frac{g_{Y}^{2}}{\sqrt{g_{3}^{2}-g_{Y}^{2}}}+Y_{3}^{F_{13}}\sqrt{g_{3}^{2}-g_{Y}^{2}},\;-Y_{12}^{F_{23}}\frac{g_{Y}^{2}}{\sqrt{g_{3}^{2}-g_{Y}^{2}}}+Y_{3}^{F_{23}}\sqrt{g_{3}^{2}-g_{Y}^{2}}\right)\,.
\end{flalign*}
Therefore, the phenomenology of $Z'_{23}$ can be completely described
in terms of its mass and the $g_{3}$ coupling. When $g_{3}$ is large,
$Z'_{23}$ is mostly coupled to the third family, while for $g_{3}$
small $Z'_{23}$ is mostly coupled to the first and second families.
In contrast, the couplings of $Z'_{12}$ need to be described in general
by two free gauge couplings. Here we choose $g_{1}$ and $g_{2}$,
but one may exchange one of these by e.g.~$g_{3}$ and $g_{Y}$ through
the matching conditions.

\begin{table}
\begin{centering}
\begin{tabular}{ll}
\toprule 
$u_{L}$ sector & $u_{R}$ sector\tabularnewline
\midrule
\midrule 
$s_{12}^{u_{L}}\simeq\frac{c_{12}^{u}}{c_{22}^{u}}\frac{\langle\phi_{q12}\rangle}{M_{U_{13}}}\sim\lambda$ & $s_{12}^{u_{R}}\simeq\frac{c_{21}^{u}}{c_{22}^{u}}\frac{\langle\phi_{\ell12}\rangle}{M_{U_{12}}}\frac{\langle\phi_{q12}\rangle}{M_{U_{13}}}\sim\lambda^{3}$\tabularnewline
\midrule 
$s_{13}^{u_{L}}\simeq\frac{c_{13}^{u}}{c_{33}^{u}}\frac{\langle\phi_{q12}\rangle}{M_{U_{13}}}\frac{\langle\phi_{q23}\rangle}{M_{U_{23}}}\sim\lambda^{3}$ & $s_{26}^{u_{R}}\simeq y_{62}^{u}\frac{\langle\phi_{\ell23}\rangle}{M_{U_{23}}}\sim\lambda^{3}$\tabularnewline
\midrule 
$s_{23}^{u_{L}}\simeq\frac{c_{23}^{u}}{c_{33}^{u}}\frac{\langle\phi_{q23}\rangle}{M_{U_{23}}}\sim\lambda^{2}$ & $s_{36}^{u_{R}}\simeq y_{63}^{u}\frac{\langle\phi_{q23}\rangle}{M_{U_{23}}}\sim\lambda^{2}$\tabularnewline
\midrule 
$s_{26}^{u_{L}}\simeq y_{26}^{u}\frac{\langle H_{3}^{u}\rangle}{M_{U_{13}}}\sim10^{-4}$ & $s_{14}^{u_{R}}\simeq y_{41}^{u}\frac{\langle\phi_{\ell12}\rangle}{M_{U_{12}}}\sim\lambda^{2}$\tabularnewline
\midrule 
$s_{15}^{u_{L}}\simeq y_{15}^{u}\frac{\langle H_{3}^{u}\rangle}{M_{U_{13}}}\sim10^{-5}$ & $s_{24}^{u_{R}}\simeq y_{42}^{u}\frac{\langle\phi_{q12}\rangle}{M_{U_{12}}}\sim\lambda$\tabularnewline
\midrule 
$s_{56}^{u_{L}}\simeq y_{56}^{u}\frac{\langle\phi_{q12}\rangle}{M_{U_{13}}}\sim\lambda$ & $s_{56}^{u_{R}}\simeq y_{65}^{u}\frac{\langle\phi_{q12}\rangle}{M_{U_{13}}}\sim\lambda$\tabularnewline
\midrule 
$s_{45}^{d_{L}}\simeq y_{45}^{d}\frac{\langle\phi_{\ell23}\rangle}{M_{U_{12,13}}}\sim\lambda^{4}$ & $s_{45}^{d_{R}}\simeq y_{54}^{d}\frac{\langle\phi_{\ell23}\rangle}{M_{U_{12,13}}}\sim\lambda^{5}$\tabularnewline
\bottomrule
\end{tabular}
\par\end{centering}
\caption{Approximate charged fermion mixing in Model 2. The angles not shown are zero
at tree-level.\label{tab:MixingModel2}}
\end{table}

The next step is to rotate the full $6\times6$ matrices containing
the gauge boson couplings to the mass basis (denoted with hat notation) for each charged sector\footnote{Although our general formalism for the gauge couplings also applies
to the neutrino sector, we do not discuss neutrino phenomenology here
since the most competitive FCNCs generated in our model involve charged
fermions only.},
\begin{equation}
\hat{\kappa}^{\psi_{L,R}}=V_{\psi_{L,R}}^{\dagger}\left(\begin{array}{cc}
\kappa_{ij}^{f_{L,R}} & 0_{3\times3}\\
0_{3\times3} & \kappa_{ab}^{F}
\end{array}\right)V_{\psi_{L,R}}\,,\quad\hat{\xi}^{\psi_{L,R}}=V_{\psi_{L,R}}^{\dagger}\left(\begin{array}{cc}
\xi_{ij}^{f_{L,R}} & 0_{3\times3}\\
0_{3\times3} & \xi_{ab}^{F}
\end{array}\right)V_{\psi_{L,R}}\,, \label{eq:FermionFullCouplings1}
\end{equation}
\begin{equation}
\hat{\omega}^{\psi_{L,R}}=V_{\psi_{L,R}}^{\dagger}\left(\begin{array}{cc}
\omega_{ij}^{f_{L,R}} & 0_{3\times3}\\
0_{3\times3} & \omega_{ab}^{F}
\end{array}\right)V_{\psi_{L,R}}\,,\label{eq:FermionFullCouplings2}
\end{equation}
where $\psi=(f,F)$ generically denotes fermions with the
same electric charge, without distinguishing between chiral and vector-like.
The mixing matrices $V_{\psi_{L,R}}$, obtained by perturbatively
diagonalising the full mass matrices of Eqs.~(\ref{eq:up_FullCouplings}-\ref{eq:e_FullCouplings}), are $6\times6$ unitary matrices that connect the interaction basis and mass basis as
\begin{equation}
    \psi_{L,R}=V_{\psi_{L,R}}\hat{\psi}_{L,R}\,.
\end{equation}
If we consider only
the mixing angles that are non-zero at tree-level, then we can write
\begin{equation}
V_{\psi_{L}}=V_{45}^{\psi_{L}}V_{56}^{\psi_{L}}V_{25}^{\psi_{L}}V_{36}^{\psi_{L}}V_{23}^{\psi_{L}}V_{13}^{\psi_{L}}V_{12}^{\psi_{L}}\,,
\end{equation}
\begin{equation}
V_{\psi_{R}}=V_{45}^{\psi_{R}}V_{56}^{\psi_{R}}V_{24}^{\psi_{R}}V_{14}^{\psi_{R}}V_{36}^{\psi_{R}}V_{26}^{\psi_{R}}V_{12}^{\psi_{R}}\,,
\end{equation}
where each $V_{\alpha\beta}^{\psi_{L,R}}$ contains a mixing angle
that quantifies the mixing between the $\alpha$ and $\beta$ families (see \cite{King:2017anf,King:2021jeo,FernandezNavarro:2022gst} for more details on the notation).
Their approximate numerical values and their scaling with parameters
of the model are given in Tables~\ref{tab:MixingModel1} and \ref{tab:MixingModel2}.

From the complete gauge coupling matrices in Eqs.~\eqref{eq:FermionFullCouplings1} and \eqref{eq:FermionFullCouplings2} we notice that
particular GIM-like mechanisms arise which protect from certain FCNCs.
For example $(f_{L1},F_{L13})$ and $(f_{L2},F_{L23})$ couple universally
to $Z'_{12}$, and similarly $(f_{L1},f_{L2})$ and $(f_{R1},f_{R2},F_{R12})$
couple universally to $Z'_{23}$. Nevertheless, significant VL-chiral
mixing in the RH sector (see Tables~\ref{tab:MixingModel1} and \ref{tab:MixingModel2}) can induce large FCNCs among
$F_{13,23}$ and light chiral fermions mediated by $Z'_{23}$,
while in the LH sector all VL-chiral mixing is suppressed by very
small angles of $\mathcal{O}(\langle H_{3}^{u,d}\rangle/M_{13,23})$.

On the other hand, we find that all right-handed fermions of each
charged sector couple universally to $Z$, i.e.~$\omega_{ij}^{f_{R}}=\omega_{ab}^{F}$,
therefore protecting from FCNCs proportional to the large RH mixing
angles. In contrast, left-handed chiral fermions and the left-handed
components of the vector-like fermions couple differently to $Z$,
i.e.~$\omega_{ij}^{f_{L}}\neq\omega_{ab}^{F}$. Giving the mixing
structure of the model, this induces 1-2 left-handed FCNCs mediated
by $Z$, which are nevertheless suppressed by small mixing angles
of $\mathcal{O}(\langle H_{3}^{u,d}\rangle/M_{13,23})$.

\subsection{Bounds from FCNCs}

The messengers of the complete models presented here are too heavy
to be directly probed at existing collider facilities. Nevertheless,
they crucially dictate the origin and size of fermion mixing which
controls the size and nature of the FCNCs mediated by the gauge bosons. 

For example, the heavy messengers (vector-like fermions) of Model
1 predict significant left-handed mixing and also significant 1-2
right-handed mixing in the three charged sectors, while the heavy
messengers of Model 2 only predict significant mixing among up-quarks
(and neutrinos). On top of this, the vector-like fermions mix with
RH chiral fermions with mildly suppressed mixing angles of $\mathcal{O}(\left\langle \phi\right\rangle /M)$,
and with LH chiral fermions with further suppressed mixing angles
of $\mathcal{O}(\langle H_{3}^{u,d}\rangle/M)$, as shown in Tables~\ref{tab:MixingModel1} and \ref{tab:MixingModel2}.
In this manner, the VL-chiral fermion mixing can induce extra flavour-violating
couplings beyond those obtained from chiral-chiral fermion mixing. 

In the following we explore the leading flavour bounds over the heavy gauge bosons and vector-like fermions of our two models, but we note that performing a complete and comprehensive study of all possible flavour-violating observables is beyond the scope of this work.
\begin{table}
\begin{centering}
\begin{tabular}{clcl}
\toprule 
\textbf{Model}  & \textbf{Observable}  & \textbf{Mediator}  & \textbf{Bound} (TeV)\tabularnewline
\midrule 
\multirow{5}{*}{\textbf{1}} & $K-\bar{K}$ (Re)  & $Z'_{12}$  & $M_{Z'_{12}}/g_{1}>340\times|\mathrm{Re}\left[\frac{c_{12}^{d}}{c_{22}^{d}}\frac{c_{21}^{d}}{c_{22}^{d}}\right]|$\tabularnewline
 & $K-\bar{K}$ (Im)  & $Z'_{12}$  & $M_{Z'_{12}}/g_{1}>3\cdot10^{3}\times|\mathrm{Im}\left[\frac{c_{12}^{d}}{c_{22}^{d}}\frac{c_{21}^{d}}{c_{22}^{d}}\right]|$\tabularnewline
\cmidrule{2-4} \cmidrule{3-4} \cmidrule{4-4} 
 & \multirow{2}{*}{$\mu\rightarrow e\gamma$ } & $Z'_{12}$  & $M_{Z'_{12}}/g_{1}>30\times|c_{12}^{e}/c_{22}^{e}|$\tabularnewline
 &  & $Z'_{23}$  & $M_{Z'_{23}}/g_{3}>8\times|y_{62}^{e}(y_{65}^{e}y_{15}^{e})^{*}|$\tabularnewline
\cmidrule{2-4} \cmidrule{3-4} \cmidrule{4-4} 
 & $\mu\rightarrow3 \, e$  & $Z'_{12}$  & $M_{Z'_{12}}/g_{1}>30\times|c_{12}^{e}/c_{22}^{e}|$\tabularnewline
\midrule 
\multirow{2}{*}{\textbf{2}} & $D-\bar{D}$ (Re)  & $Z'_{12}$  & $M_{Z'_{12}}/g_{1}>150\times|\mathrm{Re}\left[\frac{c_{12}^{u}}{c_{22}^{u}}\frac{c_{21}^{u}}{c_{22}^{u}}\right]|$\tabularnewline
 & $D-\bar{D}$ (Im)  & $Z'_{12}$  & $M_{Z'_{12}}/g_{1}>500\times|\mathrm{Im}\left[\frac{c_{12}^{u}}{c_{22}^{u}}\frac{c_{21}^{u}}{c_{22}^{u}}\right]|$\tabularnewline
\bottomrule
\end{tabular}
\par\end{centering}
\caption{Bounds in TeV on the mass of the heavy gauge bosons $M_{Z'_{12}}\sim v_{12}$
and $M_{Z'_{23}}\sim v_{23}$, assuming $g_{1}\approx g_{2}$ for
the gauge couplings. The model dependence is due to the different
fermion mixing predicted in Model 1 and 2. Re and Im refer to the
real and imaginary parts of the effective operator responsible for
a given contribution, respectively. \label{tab:Bounds}}
\end{table}
\subsubsection*{Bounds on the $\boldsymbol{Z'_{12}}$ mass}

\begin{figure}[t]
\noindent \begin{centering}
\subfloat[]{\begin{centering}
\resizebox{.46\textwidth}{!}{ \begin{tikzpicture}	
	\begin{feynman}
		\vertex (a) {\(\mu_{R}\)};
		\vertex [right=18mm of a] (b);
		\vertex [right=16mm of b] (c) ;
		\vertex [right=16mm of c] (d) ;
		\vertex [right=16mm of d] (e){\(e_{L}\)};
		\diagram* {
			(a) -- [fermion] (b) -- [boson, half left, edge label=$Z'_{12}$] (d),
			(b) -- [edge label'=\(\mu_{R}\), inner sep=6pt, insertion=1] (c),
			(c) --  [edge label'=\(\mu_{L}\), inner sep=6pt] (d),
			(d) -- [fermion] (e),
	};
	\end{feynman}
\end{tikzpicture}} 
\par\end{centering}
}$\;$\subfloat[]{\begin{centering}
\resizebox{.46\textwidth}{!}{ \begin{tikzpicture}	
	\begin{feynman}
		\vertex (a) {\(\mu_{R}\)};
		\vertex [right=18mm of a] (b);
		\vertex [right=16mm of b] (c) ;
		\vertex [right=16mm of c] (d) ;
		\vertex [right=16mm of d] (e){\(e_{L}\)};
		\diagram* {
			(a) -- [fermion] (b) -- [boson, half left, edge label=$Z'_{23}$] (d),
			(b) -- [edge label'=\(E_{23R}\), inner sep=6pt, insertion=1] (c),
			(c) --  [edge label'=\(E_{23L}\), inner sep=6pt] (d),
			(d) -- [fermion] (e),
	};
	\end{feynman}
\end{tikzpicture}} 
\par\end{centering}
}
\par\end{centering}
\caption{Leading contributions to $\mu\rightarrow e\gamma$ mediated by $Z'_{12}$
(left) and by $Z'_{23}$ (right). The crosses denote mass insertions, with the photon lines not shown.\label{fig:diagram_mu_e_gamma}}
\end{figure}
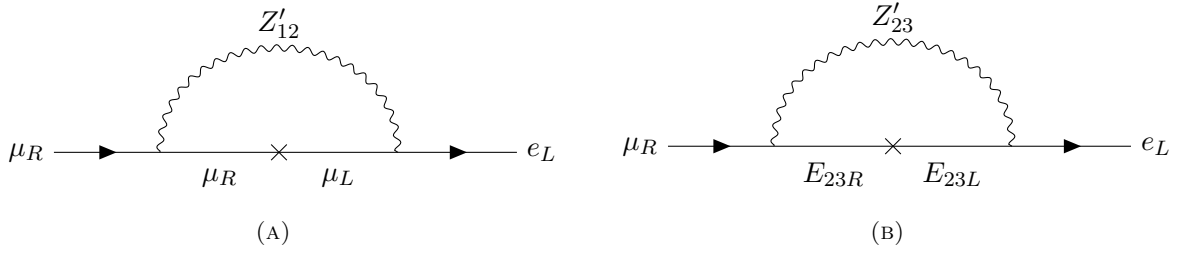

We start by discussing the phenomenology of the $Z'_{12}$ boson which
potentially mediates dangerous FCNCs, being very sensitive to the
fermion mixing dictated by the heavy messengers in both models. We
show in more detail the EFT formalism used to extract the bounds of
this section in Appendix~\ref{sec:EFT}.

In Model 1 the leading signals are expected in the down sector, and
to a lesser extent in the charged lepton sector. In particular, we
find two leading contributions to $K-\bar{K}$ mixing given by the
operators
\begin{equation}
Q_{1}^{sd}=(\bar{s}_{L}\gamma_{\mu}d_{L})^{2}\,,\qquad Q_{5}^{sd}=(\bar{s}_{L}^{\alpha}d_{R}^{\beta})(\bar{s}_{R}^{\beta}d_{L}^{\alpha})\,,
\end{equation}
generated from tree-level $Z'_{12}$ exchange (with $\alpha$ and
$\beta$ being colour indices), whose sizes are connected to the 1-2
LH and RH mixing angles $s_{12}^{d_{L}}\sim\lambda c_{12}^{d}$ and
$s_{12}^{d_{R}}\sim\lambda^{2}c_{21}^{d}$. However,
the experimental bounds over the $Q_{5}^{sd}$ operator are much stronger
than those over $Q_{1}^{sd}$.

As shown in Table~\ref{tab:Bounds}, by assuming $g_{1}\approx g_{2}$
for the gauge couplings in the multi-TeV region (as expected by the
GUT origin~\cite{FernandezNavarro:2023hrf}) and assuming all the dimensionless coefficients to be $\mathcal{O}(1)$ as expected, we find typical bounds $M_{Z'_{12}}\gtrsim\mathcal{O}(100\,\mathrm{TeV})$
from the real part of $Q_{5}^{sd}$ and $M_{Z'_{12}}\gtrsim\mathcal{O}(10^{3}\,\mathrm{TeV})$
from the imaginary part~\cite{UTfit:2007eik,Bona:2024bue,UTfit2023}, although the latter is completely dependent on the size
of complex phases in the dimensionless couplings. Then from $\mu\rightarrow e\gamma$
and $\mu\rightarrow3 \, e$ \cite{PDG:2022ynf,MEGII:2023ltw} we find the less stringent bound $M_{Z'_{12}}\gtrsim\mathcal{O}(30\,\mathrm{TeV})$ with respect to $K-\bar{K}$, due to the fact that both LH and
RH charged lepton mixing are further suppressed with respect to down-quark
mixing by extra $\lambda$ factors. In any case, the bounds are consistent
with our expectation of $M_{Z'_{12}}\sim v_{12}\sim\mathcal{O}(10^{3}\,\mathrm{TeV})$,
but current precision in $K-\bar{K}$ mixing is starting to test the
complex phases of some dimensionless coefficients in the down-quark
Yukawa matrix.

In Model 2, quark mixing is mostly generated from the up sector and
both the down-quark and charged lepton mass matrices are diagonal at tree-level.
Therefore, in this model the phenomenology of $Z'_{12}$ is dominated
by the two operators
\begin{equation}
Q_{1}^{cu}=(\bar{c}_{L}\gamma_{\mu}u_{L})^{2}\,,\qquad Q_{5}^{cu}=(\bar{c}_{L}^{\alpha}u_{R}^{\beta})(\bar{c}_{R}^{\beta}u_{L}^{\alpha})\,,
\end{equation}
that contribute to $D-\bar{D}$ mixing and are controlled by the size
of $s_{12}^{u_{L}}\sim\lambda c_{12}^{u}$ and $s_{12}^{u_{R}}\sim\lambda^{3}c_{21}^{u}$. We find the leading bound $M_{Z'_{12}}\gtrsim\mathcal{O}(100\,\mathrm{TeV})$ from the real part of $Q_{5}^{cu}$~\cite{UTfit:2007eik,Bona:2024bue,UTfit2023} and $M_{Z'_{12}}\gtrsim\mathcal{O}(500\,\mathrm{TeV})$
from the imaginary part, respectively. In both
Models 1 and 2 we find that contributions to FCNCs via VL-chiral fermion
mixing are subleading for the phenomenology of the $Z'_{12}$ boson, which is
mostly driven by the size of the 1-2 mixing angles.

\subsubsection*{Bounds on the $\boldsymbol{Z'_{23}}$ mass}

\begin{figure}[t]
\subfloat[]{
\includegraphics[scale=0.35]{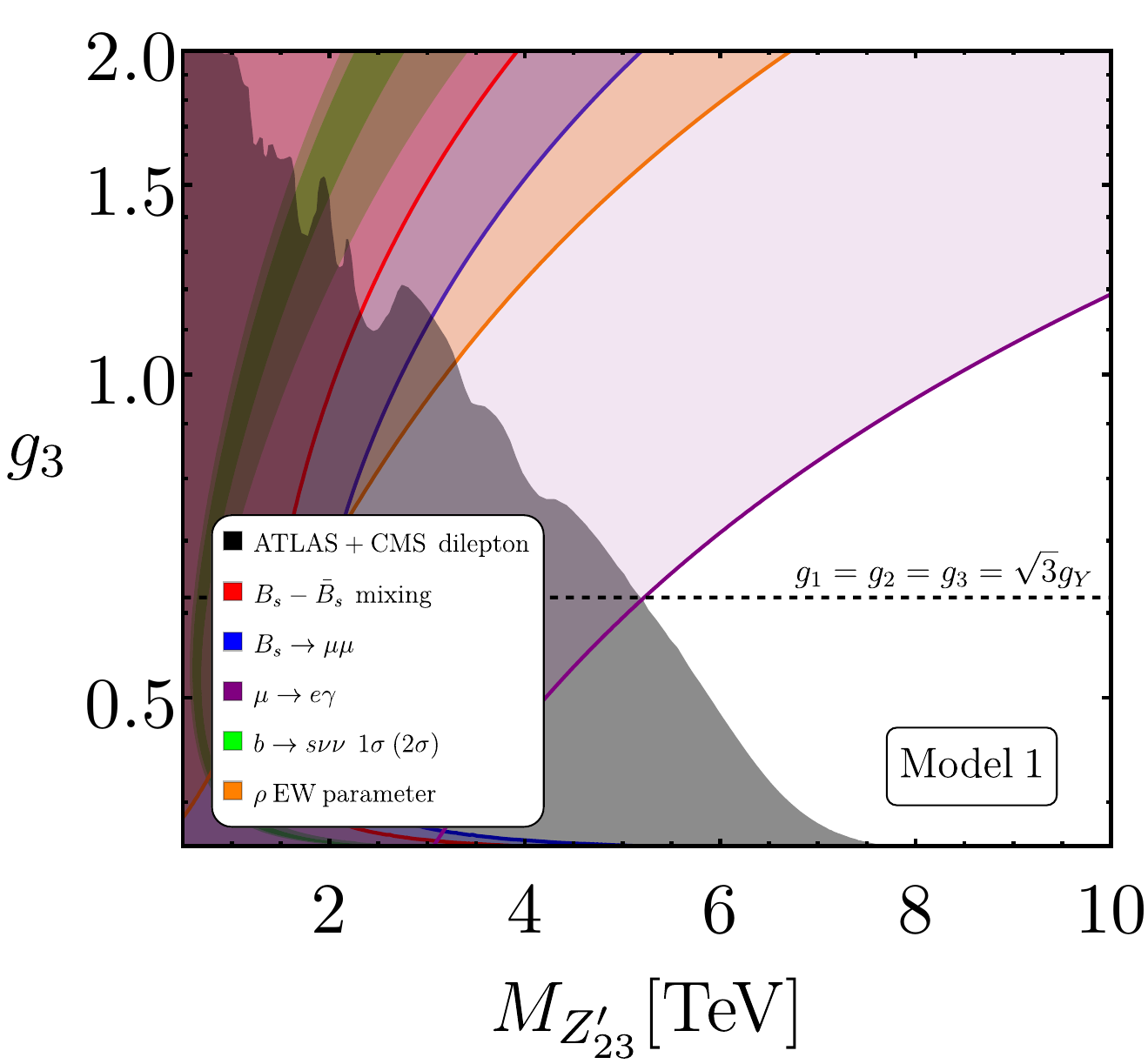}
\label{fig:Zp23a}
}$\quad$\subfloat[]{
\includegraphics[scale=0.35]{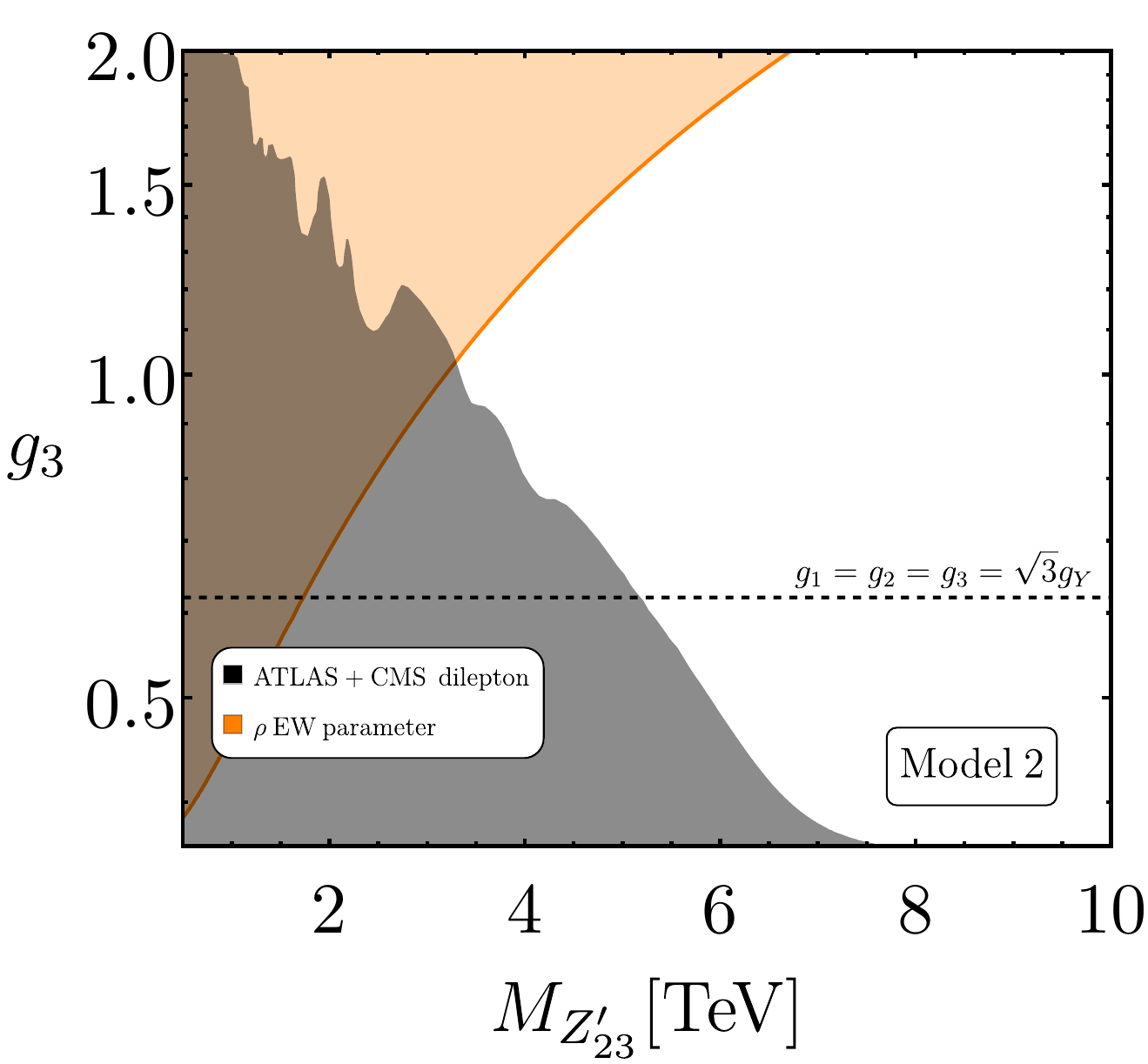}
\label{fig:Zp23b}}

\caption{Parameter space of the 23-symmetry breaking, where $M_{Z'_{23}}$ is the mass of the $Z'_{23}$ gauge boson and $g_{3}$ is the gauge coupling of the $U(1)_{Y_{3}}$ gauge group. We consider respectively the non-generic fermion mixing predicted by Model 1 (left) and by Model 2 (right) where no significant FCNCs are mediated by $Z'_{23}$, hence only the model independent observables appear in the right plot. The benchmark of equal couplings is motivated by a possible GUT origin~\cite{FernandezNavarro:2023rhv}, where the three gauge couplings are of the same order in the multi-TeV.}
\end{figure}

In contrast to $Z'_{12}$, the $Z'_{23}$ boson is mostly protected from the most
dangerous FCNCs by an approximate $U(2)^{5}$ flavour symmetry. The
exception to this appears in Model 1 only, where we find that $Z'_{23}$
mediates a 1-loop contribution to $\mu\rightarrow e\gamma$ with $E_{23}$
running in the loop, as shown in Fig.~\ref{fig:diagram_mu_e_gamma} (right), that appears due to the mixing
of the $E_{23}$ VL fermion\footnote{This is a consequence of non-minimal breaking of the approximate $U(2)^{5}$
flavour symmetry by the messengers $F_{23}$ in Model
1. This is avoided in the (less minimal) model of~\cite{FernandezNavarro:2023hrf}, where the leading
breaking of $U(2)^{5}$ is done minimally, SM-like via the VL fermion
$Q_{3}\sim(\mathbf{3,2})_{(0,0,\frac{1}{6})}$ and heavy Higgs doublets
$H_{2}^{u,d}\sim(\mathbf{1,2})_{(0,\frac{1}{2},0)}$. Alternatively,
in Model 2 the $U(2)^{5}$ symmetry is broken minimally in the down and charged
lepton sectors, and non-minimally in the up sector where experimental
bounds are much weaker. For recent revisits of $U(2)$ symmetries we refer to \cite{Greljo:2023bix,Antusch:2023shi,Greljo:2024zrj}.} with $\mu$ and $e$. This contribution to $\mu\rightarrow e\gamma$
is chirally enhanced by a factor $M_{E_{23}}/m_{\mu}\sim10^{7}$ but
also suppressed by small mixing angles $s_{26}^{e_{R}}\sim\mathcal{O}(\langle\phi_{\ell23}\rangle/M_{E_{23}}\sim\lambda^{3})$
and $s_{16}^{e_{L}}\sim\mathcal{O}(s_{12}^{e_{L}}\langle H_{3}^{d}\rangle/M_{E_{23}}\sim10^{-7})$.
As shown in Table~\ref{tab:Bounds} and Fig.~\ref{fig:Zp23a}, this contribution leads to a
bound in the few TeV region, stronger than the bounds obtained from $3\rightarrow2$ flavour transitions such as $B_{s}$ meson mixing or $B_{s}\rightarrow \mu\mu$. In  Fig.~\ref{fig:Zp23a} we also depict in green the region of parameter space that could explain the anomaly in $b\rightarrow s \nu\nu$ transitions recently observed by Belle II \cite{Belle-II:2023esi}, although this region is excluded by electroweak precision and by the leading flavour-violating observables. Similarly, we expect significant enhancements of $b\rightarrow s \mu\mu$ or $b\rightarrow s \tau\tau$ (see e.g.~\cite{Capdevila:2017iqn,Alguero:2023jeh,Bordone:2023ybl}) to be in conflict with electroweak precision and with the leading flavour observables.

This bound is well compatible with the model-independent phenomenology
of $Z'_{23}$, which is dominated by the production of dilepton tails
at the LHC and the modification of electroweak precision observables
via mixing\footnote{As shown in~\cite{FernandezNavarro:2023rhv}, in the tri-hypercharge theory there exists $Z-Z'_{23}$ mass mixing even when kinetic mixing among the three gauge $U(1)$ factors
is negligible, as it happens when they originate from a semi-simple
embedding like in~\cite{FernandezNavarro:2023hrf}. For this reason, in our analysis we neglect kinetic mixing, which is much smaller than the leading mass mixing effect that we include.} with the SM $Z$ boson~\cite{FernandezNavarro:2023rhv,Davighi:2023evx}. In particular, assuming all Yukawa couplings of the theory to be of $\mathcal{O}(1)$, and assuming $g_{3}(10\,\mathrm{TeV})\simeq0.6$ as suggested by
a GUT origin~\cite{FernandezNavarro:2023hrf}, we obtain the same bound $M_{Z'_{23}}>5\,\mathrm{TeV}$ from
both $\mu\rightarrow e\gamma$ and from
dilepton searches at the LHC.
Therefore, to be conservative we set $v_{23}\sim\mathcal{O}(10\,\mathrm{TeV})$
and, thanks to the predictive relation in Eq.~(\ref{eq:VEV_relation}),
this fixes the whole spectrum of both models as in Fig.~\ref{fig:spectrum},
and in particular the mass of $Z'_{12}$ as $M_{Z'_{12}}\sim v_{12}\sim\mathcal{O}(10^{3}\,\mathrm{TeV})$,
well compatible with all the bounds shown in Table~\ref{tab:Bounds}.

In contrast with Model 1, in Model 2 CKM mixing originates from the up sector, with the down-quark and charged lepton Yukawa couplings being diagonal at tree-level. Therefore, $Z'_{23}$ can only mediate FCNCs involving up-quarks. We find that the approximate $U(2)^{5}$ symmetry protects from contributions to $D-\overline{D}$ mixing, even in the presence of the VL quarks $U_{ij}$. Indeed only $t\rightarrow c,u$ FCNCs can be sizable, but these are poorly bounded from the experiment, leading to no significant bounds over the parameter space in Fig.~\ref{fig:Zp23b} beyond the model-independent ones that correspond to the production of dilepton tails
at the LHC and the modification of electroweak precision observables.

\subsubsection*{Bounds on vector-like fermion masses}

\begin{table}
\begin{centering}
\begin{tabular}{ccl}
\toprule 
\textbf{Order}  & \textbf{Observable}  & \textbf{Bound} (TeV)\tabularnewline
\midrule 
\multicolumn{1}{c}{} & $D-\bar{D}$ (Re)  & $M_{U_{23}}>2.5\times|\mathrm{Re}\left[y_{26}^{u}(y_{65}^{u}y_{15}^{u})^{*}\right]|$\tabularnewline
 & $D-\bar{D}$ (Im)  & $M_{U_{23}}>4\times\left|\mathrm{Im}\left[y_{26}^{u}(y_{65}^{u}y_{15}^{u})^{*}\right]\right|$\tabularnewline
\cmidrule{2-3} \cmidrule{3-3} 
Tree-level  & $K-\bar{K}$ (Re)  & $M_{D_{23}}>0.1\times|\mathrm{Re}\left[y_{26}^{d}(y_{65}^{d}y_{15}^{d})^{*}\right]|$\vspace{0.1cm}\tabularnewline
($Z$-mediated)  & $K-\bar{K}$ (Im)  & $M_{D_{23}}>0.3\times|\mathrm{Im}\left[y_{26}^{d}(y_{65}^{d}y_{15}^{d})^{*}\right]|$\tabularnewline
\cmidrule{2-3} \cmidrule{3-3} 
 & $\mu\rightarrow3 \, e$  & $M_{E_{23}}>2.2\times|y_{26}^{e}(y_{65}^{e}y_{15}^{e})^{*}|$\tabularnewline
\midrule 
\multicolumn{1}{c}{} & $D-\bar{D}$ (Re)  & $M_{U_{23}}>15\times|\mathrm{Re}\left[y_{26}^{u}(y_{65}^{u}y_{15}^{u})^{*}\right]|$\tabularnewline
1-loop  & $D-\bar{D}$ (Im)  & $M_{U_{23}}>130\times\left|\mathrm{Im}\left[y_{26}^{u}(y_{65}^{u}y_{15}^{u})^{*}\right]\right|$\tabularnewline
\cmidrule{2-3} \cmidrule{3-3} 
(box with $H_{3}^{u,d}$)  & $K-\bar{K}$ (Re)  & $M_{D_{23}}>14\times|\mathrm{Re}\left[y_{26}^{d}(y_{65}^{d}y_{15}^{d})^{*}\right]|$\tabularnewline
 & $K-\bar{K}$ (Im)  & $M_{D_{23}}>170\times|\mathrm{Im}\left[y_{26}^{d}(y_{65}^{d}y_{15}^{d})^{*}\right]|$\tabularnewline
\bottomrule
\end{tabular}
\par\end{centering}
\caption{Bounds in TeV over the mass of the heavy vector-like fermions, extracted
from FCNCs mediated by the SM $Z$ boson and from 1-loop box diagrams
involving $F_{23}$ and $H_{3}^{u,d}$. Re and Im refer to the real
and imaginary parts of the effective operator responsible for a given
contribution, respectively. Notice that $D_{23}$ and $E_{23}$ are
only present in Model 1, while $U_{23}$ is present in both Models
1 and 2. \label{tab:BoundsVLfermions}}
\end{table}

Finally, we discuss the bounds over the masses of the heavy messenger fermions obtained from FCNCs. These arise from contributions that are sensitive
to the VL fermion masses directly, rather than sensitive to ratios
of the form $\langle\phi\rangle/M_{F}$ which we consider fixed by
the flavour model. 

For starters, as shown in Section~\ref{sec:GaugeCouplings}, mixing
between vector-like and chiral fermions induces FCNCs mediated by
the SM $Z$ boson and involving left-handed fermions only. In particular,
the $Z$ boson mediates a purely left-handed contribution to $K-\bar{K}$
mixing via the effective operator $Q_{1}^{sd}$, and similarly for
$D-\bar{D}$ mixing via $Q_{1}^{cu}$. It also mediates purely left-handed
contributions to $\mu\rightarrow3 \, e$ at tree-level. These FCNCs are suppressed by small mixing angles of $\mathcal{O}(\langle H_{3}^{u}\rangle/M_{23,13}\sim10^{-5})$
in the up-sector and by $\mathcal{O}(\langle H_{3}^{d}\rangle/M_{23,13}\sim10^{-7})$
in the down and charged lepton sectors. Thanks
to the relation $M_{F_{23}}/M_{F_{12,13}}\sim\lambda$ fixed by the flavour
model, we can express all contributions in terms of the mass of the
lightest vector-like fermions, $M_{F_{23}}$, and obtain bounds over this scale.

The $Z$-mediated FCNCs are then directly sensitive to the masses
of $M_{F_{23}}$, rather than to any ratio of NP scales, providing direct bounds over the masses of heavy VL messengers
in the theory. As shown in Table~\ref{tab:BoundsVLfermions}, we
find bounds in the few TeV range over $M_{F_{23}}$,
well below their expected values of $\mathcal{O}(10^{3}\,\mathrm{TeV})$.
Particularly weak are the bounds over $M_{D_{23}}$ with respect to
those over $M_{U_{23}}$, due to the fact that the bounds over $Q_{1}^{sd}$
and $Q_{1}^{cu}$ are of similar size~\cite{UTfit:2007eik,Bona:2024bue,UTfit2023}
but the former is suppressed by the smaller VEV $\langle H_{3}^{d}\rangle$.
This is due to the fact that $\tan \beta=\langle H_{3}^{u}\rangle/\langle H_{3}^{d}\rangle\simeq\lambda^{-2}\approx20$
is imposed to explain the $m_{b,\tau}/m_{t}$ mass hierarchies, as
discussed at the beginning of Section~\ref{sec:MinimalModels}.

\begin{figure}[t]
\subfloat[]{\noindent \begin{centering}
\begin{tikzpicture}
	\begin{feynman}
		\vertex (a) {\(c_{L}\)};
		\vertex [right=24mm of a] (b);
		\vertex [right=20mm of b] (c);
		\vertex [right=20mm of c] (d) {\(u_{L}\)};
		\vertex [above=20mm of b] (f1);
		\vertex [above=20mm of c] (f2);
		\vertex [left=20mm of f1] (f3) {\(u_{L}\)};
		\vertex [right=20mm of f2] (f4) {\(c_{L}\)};
		\diagram* {
			(a) -- [anti fermion] (b) -- [scalar, edge label=\(H^{u}_{3}\)] (f1) -- [anti fermion] (f3),
			(b) -- [anti fermion, edge label'=\(U_{23R}\), inner sep=6pt] (c) -- [scalar, edge label'=\(H^{u}_{3}\)] (f2) -- [fermion] (f4),
			(c) -- [anti fermion] (d),
			(f1) -- [fermion, edge label=\(U_{23R}\), inner sep=6pt] (f2),
	};
	\end{feynman}
\end{tikzpicture}
\par\end{centering}
}$\quad$\subfloat[]{\noindent \begin{centering}
\begin{tikzpicture}
	\begin{feynman}
		\vertex (a) {\(s_{L}\)};
		\vertex [right=24mm of a] (b);
		\vertex [right=20mm of b] (c);
		\vertex [right=20mm of c] (d) {\(d_{L}\)};
		\vertex [above=20mm of b] (f1);
		\vertex [above=20mm of c] (f2);
		\vertex [left=20mm of f1] (f3) {\(d_{L}\)};
		\vertex [right=20mm of f2] (f4) {\(s_{L}\)};
		\diagram* {
			(a) -- [anti fermion] (b) -- [scalar, edge label=\(H^{d}_{3}\)] (f1) -- [anti fermion] (f3),
			(b) -- [anti fermion, edge label'=\(D_{23R}\), inner sep=6pt] (c) -- [scalar, edge label'=\(H^{d}_{3}\)] (f2) -- [fermion] (f4),
			(c) -- [anti fermion] (d),
			(f1) -- [fermion, edge label=\(D_{23R}\), inner sep=6pt] (f2),
	};
	\end{feynman}
\end{tikzpicture}
\par\end{centering}
}

\caption{1-loop box diagrams contributing to $D-\bar{D}$ mixing (left) and
to $K-\bar{K}$ mixing (right), with VL fermions $D_{23}$, $U_{23}$
and Higgs doublets $H_{3}^{u,d}$ running in the loop. \label{fig:Box_Diagrams}}
\end{figure}
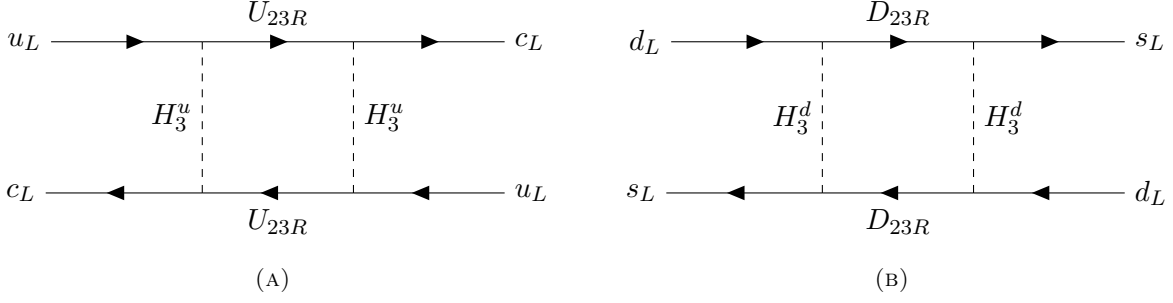

Going beyond tree-level, we find 1-loop contributions to four-fermion operators that arise from box diagrams with scalars and VL fermions running in the loop. We find the leading contributions\footnote{Diagrams involving hyperons are further suppressed by factors of $\lambda$ or involve the heavier fermions $F_{12,13}$, hence resulting in weaker bounds.} to arise from box diagrams involving the $H_{3}^{u,d}$ Higgs doublets and VL fermions $F_{23}$ running in the loop, as shown in Fig.~\ref{fig:Box_Diagrams}. These diagrams are not affected by any CKM or mass suppression but are directly sensitive to the heavy masses of the $F_{23}$ VL fermions, and to fundamental Yukawa couplings of the model expected to be of $\mathcal{O}(1)$. Box diagrams of this kind have been computed previously in the literature~\cite{Ishiwata:2015cga,Bobeth:2016llm,Alves:2023ufm}. Adapting these results to our model, assuming $M_{F_{23}}\gg M_{H_{3}^{u,d}}$, we obtain effective
contributions to $Q_{1}^{sd,cu}$ as
\begin{equation}
\left.C_{1}^{sd}\right|_{\mathrm{1-loop}}\simeq\frac{1}{8}\frac{\left(y_{26}^{d}(y_{65}^{d}y_{15}^{d})^{*}\lambda\right)^{2}}{16\pi^{2}M_{D_{23}}^{2}}\,,\qquad\left.C_{1}^{uc}\right|_{\mathrm{1-loop}}\simeq\frac{1}{8}\frac{\left(y_{26}^{u}(y_{65}^{u}y_{15}^{u})^{*}\lambda\right)^{2}}{16\pi^{2}M_{U_{23}}^{2}}\,.
\end{equation}
As shown in Table~\ref{tab:BoundsVLfermions}, from the experimental
bounds over the $Q_{1}^{sd,cu}$ operators we obtain bounds over $M_{D_{23}}$
and $M_{U_{23}}$ of $\mathcal{O}(100\,\mathrm{TeV})$, which are
compatible with our expectations of $M_{U_{23},D_{23}}\gtrsim\mathcal{O}(10^{3}\,\mathrm{TeV})$
according to the spectrum of Fig.~\ref{fig:spectrum}.

Finally, we comment that a box diagram similar to those of Fig.~\ref{fig:Box_Diagrams}
but involving charged leptons also arises in Model 1, contributing
to $\mu\rightarrow3 \, e$. However, the associated bounds turn out to be much weaker than those from meson mixing, roughly $M_{E_{23}}\apprge\mathcal{O}(1\,\mathrm{TeV})$
and hence of the same order as the bounds from tree-level $Z$-mediated
processes.

\section{Conclusions}  \label{sec:Conclusions}
Tri-hypercharge is based on assigning a separate gauge hypercharge to each fermion family. 
This simple framework has been shown to explain successfully the origin of hierarchies in the flavour structure of the SM~\cite{FernandezNavarro:2023rhv}, and may arise from a gauge unified framework~\cite{FernandezNavarro:2023hrf}. 

In this paper we have proposed and studied two minimal, ultraviolet complete, renormalisable tri-hypercharge theories of flavour. These models are the most minimal such models, in the sense of number of total degrees of freedom and representations, which can account for the quark and lepton (including neutrino) masses and mixing parameters in a natural way, namely where all observed mass and mixing hierarchies are explained in terms of ratios of high energy mass scales. These two models are similar from the low-energy point of view but differ in the heavy messenger sector that completes the theory, with consequently different flavour-changing phenomenology.

In the first model all the heavy messengers are vector-like, $SU(2)$ singlet fermions, hence containing the simplest scalar potential, and in the second model several of the vector-like fermions are replaced by just two heavy Higgs doublets, hence being more minimal in terms of total degrees of freedoms and representations but having more terms in the scalar potential.

Both models provide a very successful explanation of the hierarchical flavour structure of the SM, which is explained in terms of just three naturally small parameters defined as ratios of new physics scales (see Eq.~\eqref{eq:ThreeSmallParameters}), with all dimensionless couplings of the complete theory being of $O(1)$. Indeed, both models translate the complicated flavour structure of the SM into three simple and physical (not so) high energy scales of new physics above the electroweak scale, as shown in Fig.~\ref{fig:spectrum}, which have meaningful, non-generic phenomenological consequences which can potentially be tested experimentally. In contrast to other alternative proposals in the literature, the three NP scales of our model are completely correlated in order to explain the flavour structure of the SM, highlighting the predictivity and minimality of the framework. This means that by testing any scale of the model, one is also essentially testing the other two.

On top of this, both models are able to explain small neutrino masses and large neutrino mixing via the implementation of a simple seesaw mechanism which requires no extra scalars beyond those present in the theory to explain the charged fermion sector, being then more minimal than the mechanism suggested in~\cite{FernandezNavarro:2023rhv,FernandezNavarro:2023hrf}. Remarkably, the implementation of the seesaw mechanism in our framework does not require the presence of very small couplings nor assumes that the scales of symmetry breaking of the theory are very high. The smallness of neutrino masses is explained via very heavy singlet neutrinos, which couple to the active neutrinos thanks to messenger neutrinos that are expected to be not very heavy, perhaps in the multi-TeV to PeV region, that could be tested via their couplings to the heavy gauge bosons of the theory.

Indeed, two heavy $Z'$ bosons are predicted to live at different, hierarchical scales which are connected to the origin of fermion mass hierarchies. These heavy gauge bosons, along with the SM $Z$ boson, couple to the heavy messengers of the complete theories and predict different phenomenology for both models. 

The lighter $Z'_{23}$ boson is protected from the most dangerous FCNCs by an accidental $U(2)^{5}$ flavour symmetry, although a contribution to $\mu \rightarrow e\gamma$ arises only in Model 1 due to non-minimal breaking of the $U(2)^{5}$ symmetry by the heavy messenger fermions. However, this contribution turns out to be compatible with the model-independent phenomenology of the $Z'_{23}$ boson, which is dominated by dilepton production at the LHC and by electroweak precision observables, and allows for $M_{Z'_{23}}\sim\mathcal{O}(10\,\mathrm{TeV})$. Taking this as our benchmark mass for $Z'_{23}$ fixes all the scales of our model, and in particular the mass of the heavier gauge boson $Z'_{12}$ as $M_{Z'_{12}}\sim \mathcal{O}(10^{3}\,\mathrm{TeV})$.

In Model 1, the heavier $Z'_{12}$ boson mediates contributions to neutral kaon mixing, $\mu \rightarrow e\gamma$ and $\mu \rightarrow 3 \, e$ at acceptable rates for the expected mass $M_{Z'_{12}}\gtrsim \mathcal{O}(10^{3}\,\mathrm{TeV})$. In contrast, in Model 2, $Z'_{12}$ only mediates contributions to neutral $D$-meson mixing which suggests $M_{Z'_{12}}\gtrsim \mathcal{O}(100\,\mathrm{TeV})$. Finally, the vector-like messenger fermions also generate FCNCs via box diagrams involving the light Higgs doublets, and induce tree-level FCNCs mediated by the SM $Z$ boson. In either case, the resulting bounds are compatible with the heavy masses expected for the messenger fermions.

In conclusion, the minimal models presented here highlight the simplicity and effectiveness of the tri-hypercharge framework to explain the origin of fermion masses and mixings, and they provide complete but simple setups to investigate the potential experimental signals associated to this class of models. On top of the model-independent phenomenology associated to direct collider searches and electroweak precision observables, the models proposed here predict complementary signals in flavour-violating observables that allow to test the higher scales of the theory, and highlight the potential of the flavour precision program to test the origin of flavour in the future.

\section*{Acknowledgements}
MFN is grateful to Xavier Ponce D\'iaz and Admir Greljo for interesting discussions about flavour model-building during the La Thuile 2024 conference.
SFK would like to thank CERN for hospitality and acknowledges the STFC Consolidated Grant ST/L000296/1 and the European Union's Horizon 2020 Research and Innovation programme under Marie Sklodowska-Curie grant agreement HIDDeN European ITN project (H2020-MSCA-ITN-2019//860881-HIDDeN). MFN is supported by the STFC under grant ST/X000605/1. AV acknowledges financial support from the Spanish grants PID2020-113775GB-I00 (AEI/10.13039/ 501100011033) and CIPROM/2021/054 (Generalitat Valenciana), as well as from MINECO through the Ram\'on y Cajal contract RYC2018-025795-I.

\appendix

\section{Complete scalar potential\label{app:ScalarPotential}}

We introduce a $\mathbb{Z}_{2}$ symmetry which enforces that
the Higgs doublets only couple respectively to up quarks (neutrinos) and down
quarks (charged leptons), as the notation of this paper suggests. This is achieved by assuming that the following fields are odd under $\mathbb{Z}_{2}$,
\begin{equation}
H_{3}^{u}\rightarrow-H_{3}^{u}\,,
\end{equation}
\begin{equation}
u_{i}^{c}\rightarrow-u_{i}^{c}\,,\qquad U_{ij}\,(\overline{U}_{ij})\rightarrow-U_{ij}\,(\overline{U}_{ij})\,,\qquad\nu_{i}^{c}\rightarrow-\nu_{i}^{c}\,,\qquad N_{ij}\,(\overline{N}_{ij})\rightarrow-N_{ij}\,(\overline{N}_{ij})\,,
\end{equation}
while all the remaining fields of both Model 1 and 2 remain even. Notice
that such $\mathbb{Z}_{2}$ should be softly broken in order to avoid
the formation of domain walls in the early Universe, hence we allow
dimensionful terms breaking $\mathbb{Z}_{2}$ explicitly in the scalar
potential, which is given in Model 1 by
\begin{flalign}
V_{1} & =m_{q_{12}}^{2}\left|\phi_{q_{12}}\right|^{2}+m_{q_{23}}^{2}\left|\phi_{q_{23}}\right|^{2}+m_{\ell12}^{2}\left|\phi_{\ell_{12}}\right|^{2}+m_{\ell23}^{2}\left|\phi_{\ell_{23}}\right|^{2} \nonumber\\
 & +\lambda_{q_{12}}\left|\phi_{q_{12}}\right|^{4}+\lambda_{q_{23}}\left|\phi_{q_{23}}\right|^{4}+\lambda_{\ell_{12}}\left|\phi_{\ell_{12}}\right|^{4}+\lambda_{\ell_{23}}\left|\phi_{\ell_{23}}\right|^{4} \nonumber\\
 & +\left(\lambda_{q_{12}^{3}\ell_{12}}\phi_{q_{12}}^{3}\phi_{\ell_{12}}+\lambda_{q_{23}^{3}\ell_{23}}\phi_{q_{23}}^{3}\phi_{\ell_{23}}+\mathrm{h.c.}\right) \nonumber\\
 & +\lambda_{q_{12}q_{23}}\left|\phi_{q_{12}}\right|^{2}\left|\phi_{q_{23}}\right|^{2}+\lambda_{q_{12}\ell_{12}}\left|\phi_{q_{12}}\right|^{2}\left|\phi_{\ell_{12}}\right|^{2}+\lambda_{q_{12}\ell_{23}}\left|\phi_{q_{12}}\right|^{2}\left|\phi_{\ell_{23}}\right|^{2} \nonumber\\
 & +\lambda_{q_{23}\ell_{12}}\left|\phi_{q_{23}}\right|^{2}\left|\phi_{\ell_{12}}\right|^{2}+\lambda_{q_{23}\ell_{23}}\left|\phi_{q_{23}}\right|^{2}\left|\phi_{\ell_{23}}\right|^{2}+\lambda_{\ell_{12}\ell_{23}}\left|\phi_{\ell_{12}}\right|^{2}\left|\phi_{\ell_{23}}\right|^{2} \nonumber\\
 & +M_{H_{3}^{u}}^{2}|H_{3}^{u}|^{2}+M_{H_{3}^{d}}^{2}|H_{3}^{d}|^{2}+\left(M_{H_{3}^{u,d}}^{2}H_{3}^{u}H_{3}^{d}+\mathrm{h.c.}\right)+\lambda_{H_{3}^{u}}|H_{3}^{u}|^{4}+\lambda_{H_{3}^{d}}|H_{3}^{d}|^{4} \nonumber\\
 & +\lambda_{H_{3}^{u}H_{3}^{d}}|H_{3}^{u}|^{2}|H_{3}^{d}|^{2}+\lambda_{\widetilde{H}_{3}^{u}\widetilde{H}_{3}^{d}}(H_{3}^{u}H_{3}^{d})(\widetilde{H}_{3}^{d}\widetilde{H}_{3}^{u})+\left(\lambda_{H_{3}^{u}H_{3}^{u}H_{3}^{d}H_{3}^{d}}H_{3}^{u}H_{3}^{u}H_{3}^{d}H_{3}^{d}+\mathrm{h.c.}\right) \nonumber\\
 & +\lambda_{H_{3}^{u}q_{12}}|H_{3}^{u}|^{2}\left|\phi_{q_{12}}\right|^{2}+\lambda_{H_{3}^{u}q_{23}}|H_{3}^{u}|^{2}\left|\phi_{q_{23}}\right|^{2}+\lambda_{H_{3}^{u}\ell_{12}}|H_{3}^{u}|^{2}\left|\phi_{\ell_{12}}\right|^{2}+\lambda_{H_{3}^{u}\ell_{23}}|H_{3}^{u}|^{2}\left|\phi_{\ell_{23}}\right|^{2} \nonumber\\
 & +\lambda_{H_{3}^{d}q_{12}}|H_{3}^{d}|^{2}\left|\phi_{q_{12}}\right|^{2}+\lambda_{H_{3}^{d}q_{23}}|H_{3}^{d}|^{2}\left|\phi_{q_{23}}\right|^{2}+\lambda_{H_{3}^{d}\ell_{12}}|H_{3}^{d}|^{2}\left|\phi_{\ell_{12}}\right|^{2}+\lambda_{H_{3}^{d}\ell_{23}}|H_{3}^{d}|^{2}\left|\phi_{\ell_{23}}\right|^{2}\,.
\end{flalign}
The scalar potential of Model 2 contains all the terms of Model 1
plus a few more involving the Higgs doublets $H_{1}^{d}$ and $H_{2}^{d}$,
\begin{flalign}
V_{2} & =V_{1}+\sum_{i=1,2}\left(M_{H_{i}^{d}}^{2}|H_{i}^{d}|^{2}+\lambda_{H_{i}^{d}}|H_{i}^{d}|^{4}+\lambda_{H_{3}^{u}H_{i}^{d}}|H_{3}^{u}|^{2}|H_{i}^{d}|^{2}+\lambda_{H_{3}^{d}H_{i}^{d}}|H_{3}^{d}|^{2}|H_{i}^{d}|^{2}\right. \nonumber\\
 & +\lambda_{\widetilde{H}_{3}^{u}\widetilde{H}_{i}^{d}}(H_{3}^{u}H_{i}^{d})(\widetilde{H}_{i}^{d}\widetilde{H}_{3}^{u})+\lambda_{\widetilde{H}_{3}^{d}\widetilde{H}_{i}^{d}}(\widetilde{H}_{3}^{d}H_{i}^{d})(\widetilde{H}_{i}^{d}H_{3}^{d}) \nonumber\\
 & \left.+\lambda_{H_{i}^{d}q_{12}}|H_{i}^{d}|^{2}\left|\phi_{q_{12}}\right|^{2}+\lambda_{H_{i}^{d}q_{23}}|H_{i}^{d}|^{2}\left|\phi_{q_{23}}\right|^{2}+\lambda_{H_{i}^{d}\ell_{12}}|H_{i}^{d}|^{2}\left|\phi_{\ell_{12}}\right|^{2}+\lambda_{H_{i}^{d}\ell_{23}}|H_{i}^{d}|^{2}\left|\phi_{\ell_{23}}\right|^{2}\right) \nonumber\\
 & +\lambda_{H_{1}^{d}H_{2}^{d}}|H_{1}^{d}|^{2}|H_{2}^{d}|^{2}+\lambda_{\widetilde{H}_{1}^{d}\widetilde{H}_{2}^{d}}(\widetilde{H}_{1}^{d}H_{2}^{d})(\widetilde{H}_{2}^{d}H_{1}^{d}) \nonumber\\
 & +\left(f_{12}^{dd}\widetilde{H}_{1}^{d}H_{2}^{d}\widetilde{\phi}_{\ell12}+f_{23}^{dd}\widetilde{H}_{2}^{d}H_{3}^{d}\widetilde{\phi}_{\ell23}+f_{23}^{du}H_{2}^{d}H_{3}^{u}\phi_{\ell23}+\lambda_{H_{1}^{d}H_{2}^{d}\phi_{12}\phi_{23}}\widetilde{H}_{1}^{d}H_{3}^{d}\widetilde{\phi}_{\ell12}\widetilde{\phi}_{\ell23}+\mathrm{h.c.}\right)\ 
\end{flalign}
Notice that the $d=4$ couplings $\phi_{q_{12}}^{3}\phi_{\ell_{12}}$
and $\phi_{q_{23}}^{3}\phi_{\ell_{23}}$ break the global phases associated
to the four hyperons, while the $d=4$ coupling $H_{3}^{u}H_{3}^{u}H_{3}^{d}H_{3}^{d}$
breaks the global phases associated to the third family Higgs doublets
(along with the $d=2$ coupling $H_{3}^{u}H_{3}^{d}$). For Model
2, the same discussion applies but the extra $d=4$ coupling $\widetilde{H}_{1}^{d}H_{3}^{d}\widetilde{\phi}_{\ell12}\widetilde{\phi}_{\ell23}$
and $d=3$ couplings $f_{12,23}^{dd}$, $f_{23}^{du}$ break the global
phases of $H_{1}^{d}$ and $H_{2}^{d}$, leaving the three gauge hypercharges
as the only remaining unbroken $U(1)$s. In this manner, both Model
1 and Model 2 are free from the appearance of physical Nambu-Goldstone
bosons.

\pagebreak[2]

Having discussed the scalar sector of our two models, we comment that we expect radiative corrections to disturb the spectrum of scales depicted in Fig.~\ref{fig:spectrum} unless some of the couplings of the scalar potential are fine-tuned to some degree. Indeed the light Higgs doublets $H^{u,d}_{3}$ that perform electroweak symmetry breaking couple to heavy degrees of freedom at tree-level, especially in Model 2 where we find that $H^{d}_{3}$ couples to the heavy Higgs doublets and the hyperons. This is a consequence of the well-known \textit{hierarchy problem}. Then the radiative stability of the scales in Fig.~\ref{fig:spectrum} requires fine-tuning of the couplings of the scalar potential to some degree, as discussed e.g.~in \cite{Allwicher:2020esa} for a related framework. In order to avoid such fine-tuning one would need to impose extra dynamics like Supersymmetry or a strongly coupled sector in order to protect the masses of fundamental scalars from radiative corrections, which is beyond the scope of this paper.

\section{Models with less than four hyperons} \label{app:ModelsLessHyperons}

In both minimal models presented in the main text, we have considered
only the four hyperons $\phi_{q12}^{(-\frac{1}{6},\frac{1}{6},0)}$,
$\phi_{\ell12}^{(\frac{1}{2},-\frac{1}{2},0)}$, $\phi_{q23}^{(0,-\frac{1}{6},\frac{1}{6})}$
and $\phi_{\ell23}^{(0,\frac{1}{2},-\frac{1}{2})}$. However, as first
shown in \cite{FernandezNavarro:2023rhv}, a simple spurion analysis reveals that one can exchange
the $\ell$-type hyperons by cubic powers of the $q$-type hyperons,
i.e.
\begin{equation}
\phi_{\ell12}\sim\tilde{\phi}_{q12}^{3}\,,\label{eq:Spurionic1}
\end{equation}
\begin{equation}
\phi_{\ell23}\sim\tilde{\phi}_{q23}^{3}\,.\label{eq:Spurionic2}
\end{equation}
Therefore, in principle it is possible to remove $\phi_{\ell12}$
and $\phi_{\ell23}$ since $\phi_{q12}$ and $\phi_{q23}$ suffice
to spontaneously break the tri-hypercharge symmetry in the desired
way of Eq.~\eqref{eq:SymmetryBreaking}, and also can generate all the spurions needed to populate the
effective Yukawa matrices. Assuming UV completions where all the heavy
messengers are vector-like fermions (like in Model 1 of the main text),
the spurionic relations in Eqs.~\eqref{eq:Spurionic1} and \eqref{eq:Spurionic2} lead to models that contain only
the two hyperons $\phi_{q12}$ and $\phi_{q23}$ and the two Higgs
doublets $H^{u,d}_{3}$ as the only scalars of the theory, hence having the most
economical scalar sector and the simplest scalar potential (and being automatically
free from the appearance of physical Goldstone bosons). 

Indeed by using only $\phi_{q12}$ and $\phi_{q23}$ we can write
the effective Yukawa matrices in an EFT framework as
\begin{flalign}
{\cal L} & =\begin{pmatrix}Q_{1} & Q_{2} & Q_{3}\end{pmatrix}\begin{pmatrix}\tilde{\phi}_{q12}^{3}\tilde{\phi}_{q23}^{3} & {\displaystyle \phi_{q12}\tilde{\phi}_{q23}^{3}} & {\phi}_{q12}{\phi}_{q23}\\
\tilde{\phi}_{q12}^{4}\tilde{\phi}_{q23}^{3} & \tilde{\phi}_{q23}^{3} & {\phi}_{q23}\\
\tilde{\phi}_{q12}^{4}\tilde{\phi}_{q23}^{4} & \tilde{\phi}_{q23}^{3}\tilde{\phi}_{q23} & 1
\end{pmatrix}\begin{pmatrix}u_{1}^{c}\\
u_{2}^{c}\\
u_{3}^{c}
\end{pmatrix}H_{3}^{u}\label{eq:eff_Yukawa_up2-1}\\
 & +\begin{pmatrix}Q_{1} & Q_{2} & Q_{3}\end{pmatrix}\begin{pmatrix}\phi_{q12}^{3}\phi_{q23}^{3} & {\displaystyle {\phi}_{q12}\phi_{q23}^{3}} & {\displaystyle {\phi}_{q12}{\phi}_{q23}}\\
{\phi}_{q12}^{2}\phi_{q23}^{3} & \phi_{q23}^{3} & {\phi}_{q23}\\
{\phi}_{q12}^{2}{\phi}_{q23}^{2} & {\phi}_{q23}^{2} & 1
\end{pmatrix}\begin{pmatrix}d_{1}^{c}\\
d_{2}^{c}\\
d_{3}^{c}
\end{pmatrix}H_{3}^{d}\\
 & +\begin{pmatrix}L_{1} & L_{2} & L_{3}\end{pmatrix}\begin{pmatrix}\phi_{q12}^{3}\phi_{q23}^{3} & \tilde{\phi}_{q12}^{3}\phi_{q23}^{3} & \tilde{\phi}_{q12}^{3}\tilde{\phi}_{q23}^{3}\\
\phi_{q12}^{6}\phi_{q23}^{3} & \phi_{q23}^{3} & \tilde{\phi}_{q23}^{3}\\
\phi_{q12}^{6}\phi_{q23}^{6} & \phi_{q23}^{6} & 1
\end{pmatrix}\begin{pmatrix}e_{1}^{c}\\
e_{2}^{c}\\
e_{3}^{c}
\end{pmatrix}H_{3}^{d}\,+\mathrm{h.c.}\,,
\end{flalign}
where the cut-off of the EFTs are not shown, we assume that a power
of $\Lambda_{1}^{-1}$ accompanies every insertion of $\phi_{q12}$
and a power of $\Lambda_{2}^{-1}$ accompanies every insertion of
$\phi_{q23}$. From the inspection of the matrices above, we already
notice that the 23-mass hierarchy $m_{2}/m_{3}\sim\phi_{q23}^{3}/\Lambda_{2}^{3}$
is generated at dimension-7 in the EFT while $V_{cb}\sim\phi_{q23}/\Lambda_{2}$,
which is numerically close, is generated at only dimension-5. Nevertheless,
after exchanging the hyperons by their VEVs, we can assume the following
numerical values,
\begin{equation}
\frac{\langle{\phi}_{q12}\rangle}{\Lambda_{1}}\sim{\displaystyle \frac{\langle{\phi}_{q23}\rangle}{\Lambda_{2}}}\sim\lambda\,,\label{eq:TwoHyperonRatios}
\end{equation}
to obtain the approximate textures (up to $\mathcal{O}(1)$ coefficients)
\begin{flalign}
{\cal L} & =\begin{pmatrix}u_{1} & u_{2} & u_{3}\end{pmatrix}\begin{pmatrix}\lambda^{6} & \lambda^{4} & \lambda^{2}\\
\lambda^{7} & \lambda^{3} & \lambda\\
\lambda^{9} & \lambda^{5} & 1
\end{pmatrix}\begin{pmatrix}u_{1}^{c}\\
u_{2}^{c}\\
u_{3}^{c}
\end{pmatrix}\frac{v_{\mathrm{SM}}}{\sqrt{2}}\\
 & +\begin{pmatrix}d_{1} & d_{2} & d_{3}\end{pmatrix}\begin{pmatrix}\lambda^{6} & \lambda^{4} & \lambda^{2}\\
\lambda^{5} & \lambda^{3} & \lambda\\
\lambda^{6} & \lambda^{4} & 1
\end{pmatrix}\begin{pmatrix}d_{1}^{c}\\
d_{2}^{c}\\
d_{3}^{c}
\end{pmatrix}\lambda^{2}\frac{v_{\mathrm{SM}}}{\sqrt{2}}\\
 & +\begin{pmatrix}e_{1} & e_{2} & e_{3}\end{pmatrix}\begin{pmatrix}\lambda^{6} & \lambda^{6} & \lambda^{6}\\
\lambda^{9} & \lambda^{3} & \lambda^{3}\\
\lambda^{12} & \lambda^{6} & 1
\end{pmatrix}\begin{pmatrix}e_{1}^{c}\\
e_{2}^{c}\\
e_{3}^{c}
\end{pmatrix}\lambda^{2}\frac{v_{\mathrm{SM}}}{\sqrt{2}}\,+\mathrm{h.c.}\,.
\end{flalign}
The textures above are successful to explain the flavour structure
of the SM, albeit requiring some dimensionless coefficients of $\mathcal{O}(\lambda)$
to generate the right $V_{cb}$ and $V_{ub}$. We consider this acceptable,
and perhaps could be avoided if the mass of a VL quark in the UV theory
provides an extra $\lambda$ factor beyond the simplified assumption
of Eq.~\eqref{eq:TwoHyperonRatios}.

\begin{table}
\begin{centering}
\begin{tabular}{ccccc}
\toprule 
 & $U(1)_{Y_{1}}$  & $U(1)_{Y_{2}}$  & $U(1)_{Y_{3}}$  & $SU(3)_{c}\times SU(2)_{L}$\tabularnewline
\midrule
$H_{3}^{u,d}$  & 0  & 0  & $\pm1/2$  & $(\mathbf{1,2})$\tabularnewline
\midrule 
$\phi_{q_{12}}$  & -1/6  & 1/6  & 0  & $(\mathbf{1,1})$\tabularnewline
$\phi_{q_{23}}$  & 0  & -1/6  & 1/6  & $(\mathbf{1,1})$\tabularnewline
\midrule 
\rowcolor{yellow!10}$U_{23}$  & 0  & -1/6  & -1/2  & $(\mathbf{\overline{3},1})$\tabularnewline
\rowcolor{yellow!10}$\widetilde{U}_{23}$  & 0  & -1/3  & --1/3  & $(\mathbf{\overline{3},1})$\tabularnewline
\rowcolor{yellow!10}$\hat{U}_{23}$  & 0  & -1/2  & -1/6  & $(\mathbf{\overline{3},1})$\tabularnewline
\rowcolor{yellow!10}$U_{13}$  & -1/6  & 0  & -1/2  & $(\mathbf{\overline{3},1})$\tabularnewline
\rowcolor{yellow!10}$U_{2}$  & 0  & 2/3  & 0  & $(\mathbf{\overline{3},1})$\tabularnewline
\rowcolor{yellow!10}$U_{12}$  & -1/6  & 1/2  & 0  & $(\mathbf{\overline{3},1})$\tabularnewline
\rowcolor{yellow!10}$\widetilde{U}_{12}$  & -1/3  & -1/3  & 0  & $(\mathbf{\overline{3},1})$\tabularnewline
\rowcolor{yellow!10}$\hat{U}_{12}$  & -1/2  & 1/6  & 0  & $(\mathbf{\overline{3},1})$\tabularnewline
\midrule
\rowcolor{yellow!10}$D_{23}$  & 0  & -1/6  & 1/2  & $(\mathbf{\overline{3},1})$\tabularnewline
\rowcolor{yellow!10}$D_{3}$  & 0  & 0  & 1/3  & $(\mathbf{\overline{3},1})$\tabularnewline
\rowcolor{yellow!10}$\tilde{D}_{23}$  & 0  & 1/6  & 1/6  & $(\mathbf{\overline{3},1})$\tabularnewline
\rowcolor{yellow!10}$D_{13}$  & -1/6  & 0  & 1/2  & $(\mathbf{\overline{3},1})$\tabularnewline
\rowcolor{yellow!10}$D_{2}$  & 0  & 1/3  & 0  & $(\mathbf{\overline{3},1})$\tabularnewline
\rowcolor{yellow!10}$\tilde{D}_{12}$  & 1/6  & 1/6  & 0  & $(\mathbf{\overline{3},1})$\tabularnewline
\midrule
\rowcolor{yellow!10}$E_{23}$  & 0  & 1/2  & 1/2  & $(\mathbf{1,1})$\tabularnewline
\rowcolor{yellow!10}$\widetilde{E}_{23}$  & 0  & 2/3  & 1/3  & $(\mathbf{1,1})$\tabularnewline
\rowcolor{yellow!10}$\hat{E}_{23}$  & 0  & 5/6  & 1/6  & $(\mathbf{1,1})$\tabularnewline
\rowcolor{yellow!10}$E_{12}$  & 1/2  & 1/2  & 0  & $(\mathbf{1,1})$\tabularnewline
\rowcolor{yellow!10}$\widetilde{E}_{12}$  & 2/3  & 1/3  & 0  & $(\mathbf{1,1})$\tabularnewline
\rowcolor{yellow!10}$\hat{E}_{12}$  & 5/6  & 1/6  & 0  & $(\mathbf{1,1})$\tabularnewline
\rowcolor{yellow!10}$E_{13}$  & 1/2  & 0  & 1/2  & $(\mathbf{1,1})$\tabularnewline
\rowcolor{yellow!10}$E_{123}$  & 1/2  & 1/3  & 1/6  & $(\mathbf{1,1})$\tabularnewline
\rowcolor{yellow!10}$\widetilde{E}_{123}$  & 1/2  & 1/6  & 1/3  & $(\mathbf{1,1})$\tabularnewline
\bottomrule
\end{tabular}
\par\end{centering}
\caption{Scalar and vector-like content of the renormalisable model with two hyperons (charged fermion sector only). As in Tables~\ref{tab:UVmodel1} and \ref{tab:UVmodel2}, for vector-like fermions (highlighted in yellow),
the conjugate partners are also included but not explicitly shown. \label{tab:Model2Hyperons}}
\end{table}

However, we find that in order to UV-complete the effective theory
via $SU(2)_{L}$ singlet VL fermions, a total of 23 VL fermion representations
are needed for the charged fermion sector only, as shown in Table~\ref{tab:Model2Hyperons}, in contrast with only
9 needed in Model 1 of the main text. Moreover, some of the VL fermions
need to be charged under the three family hypercharges. Hence we find
that in models with only two hyperons $\phi_{q12}$ and $\phi_{q23}$,
despite containing the simplest scalar sector and the most economical
low-energy theory, the UV-complete theories seem to be far from minimal.

In comparison with Model 1 of the main text, the model sketched here
contains two less hyperons but at least 14 more VL fermions. This counting
is obtained without exploring the neutrino sector, which will require the addition
of more neutrino messengers and the full neutrino mass matrix will
potentially contain various powers of the VEVs $\langle\phi_{q12}\rangle$
and $\langle\phi_{q23}\rangle$, perhaps making it more difficult
to explain large neutrino mixing. Instead, if we seek for UV completions
as in Model 2 of the main text, we will need to introduce 9 heavy
Higgs doublets for the down and charged lepton sector (plus 7 VL
fermions for the up sector), hence being less minimal and also complicating
the scalar potential.

Perhaps an alternative would be to introduce $\phi_{\ell23}^{(0,\frac{1}{2},-\frac{1}{2})}$
only, without introducing $\phi_{\ell12}^{(\frac{1}{2},-\frac{1}{2},0)}$.
This is exactly the model with 3 hyperons first discussed in Section~3.3
of \cite{FernandezNavarro:2023rhv}. In this case, the effective Yukawa matrix (showing only the
up sector for simplicity) reads
\begin{equation}
{\cal L}_{u}=\begin{pmatrix}Q_{1} & Q_{2} & Q_{3}\end{pmatrix}\begin{pmatrix}\tilde{\phi}_{q12}^{3}{\phi}_{\ell23} & {\phi}_{q12}{\phi}_{\ell23} & {\phi}_{q12}{\phi}_{q23}\\
\tilde{\phi}_{q12}^{4}{\phi}_{\ell23} & {\phi}_{\ell23} & {\phi}_{q23}\\
\tilde{\phi}_{q12}^{4}{\phi}_{\ell23}\tilde{\phi}_{q23} & {\phi}_{\ell23}\tilde{\phi}_{q23} & 1
\end{pmatrix}\begin{pmatrix}u_{1}^{c}\\
u_{2}^{c}\\
u_{3}^{c}
\end{pmatrix}H^{u}_{3}+\mathrm{h.c.}
\end{equation}
hence now both $m_{2}/m_{3}$ and $V_{cb}$ are generated at dimension-5
in the EFT expansion, but in contrast with the models of the main
text now we generate first family masses at dimension-8 rather than
dimension-6, due to the fact that $\phi_{\ell12}^{(\frac{1}{2},-\frac{1}{2},0)}$
is not present but generated as $\phi_{\ell12}\sim\phi_{q12}^{3}$.
As first noted in \cite{FernandezNavarro:2023rhv}, from the pure EFT point of view, this provides
a nice description of the SM flavour hierarchies thanks to further hyperon
insertions for the smallest (effective) Yukawa couplings.

However, we find that in order to UV-complete this effective theory
via $SU(2)_{L}$ singlet VL fermions, a total of 15 VL fermion representations
are needed for the charged fermion sector only, in contrast with only
9 needed in Model 1 of the main text. Therefore, in comparison with
Model 1 of the main text, this model would contain only one less hyperon
but 6 more VL fermions. On top of this, in this model the hyperons
$\phi_{q23}$ and $\phi_{\ell23}$ have a quartic coupling in the scalar potential of the form $\phi_{q23}^{3} \, \phi_{\ell23}$. This breaks the global phases
of both hyperons. In contrast, $\phi_{q12}$ cannot couple to other scalars beyond quadratic self-conjugate terms, hence a physical Goldstone boson potentially arises associated to the unbroken phase of $\phi_{q12}$. This problem persists in UV completions based on the addition of heavy Higgs doublets, as in Model 2 of the
main text, where we find that at least 5 heavy Higgs doublets are
needed for the down and charged lepton sectors.

\newpage

\section{EFT formalism for FCNCs} \label{sec:EFT}

The leading FCNCs in the quark sector are connected to the following
operators contributing to $K-\bar{K}$ and $D-\overline{D}$ mixing
\begin{equation}
\mathcal{L}_{\mathrm{EFT}}\supset-\left[C_{LL}^{sd}(\bar{s}_{L}\gamma_{\mu}d_{L})^{2}+C_{LR}^{sd}(\bar{s}_{L}\gamma_{\mu}d_{L})(\bar{s}_{R}\gamma_{\mu}d_{R})+C_{LL}^{cu}(\bar{c}_{L}\gamma_{\mu}u_{L})^{2}+C_{LR}^{cu}(\bar{c}_{L}\gamma_{\mu}u_{L})(\bar{c}_{R}\gamma_{\mu}u_{R})\right]\,,
\end{equation}
where
\begin{flalign}
C_{LL}^{sd,uc} & =\frac{\hat{\kappa}_{12}^{d_{L},u_{L}}\hat{\kappa}_{12}^{d_{L},u_{L}}}{2M_{Z'_{12}}^{2}}+\frac{\hat{\xi}_{12}^{d_{L},u_{L}}\hat{\xi}_{12}^{d_{L},u_{L}}}{2M_{Z'_{23}}^{2}}+\frac{\hat{\omega}_{12}^{d_{L},u_{L}}\hat{\omega}_{12}^{d_{L},u_{L}}}{2M_{Z}^{2}}\,,\\
C_{LR}^{sd,uc} & =\frac{\hat{\kappa}_{12}^{d_{L},u_{L}}\hat{\kappa}_{12}^{d_{R},u_{R}}}{M_{Z'_{12}}^{2}}+\frac{\hat{\xi}_{12}^{d_{L},u_{L}}\hat{\xi}_{12}^{d_{R},u_{R}}}{M_{Z'_{23}}^{2}}\,.
\end{flalign}
Now we apply a Fierz transformation to the LR operators in order to
match to the basis of Refs.~\cite{UTfit:2007eik,Bona:2024bue,UTfit2023}, from which we take the bounds,
\begin{equation}
\mathcal{L}_{\mathrm{EFT}}\supset-\left[C_{1}^{sd}(\bar{s}_{L}\gamma_{\mu}d_{L})^{2}+C_{5}^{sd}(\bar{s}_{L}^{\alpha}d_{R}^{\beta})(\bar{s}_{R}^{\beta}d_{L}^{\alpha})+C_{1}^{cu}(\bar{c}_{L}\gamma_{\mu}u_{L})^{2}+C_{5}^{cu}(\bar{c}_{L}^{\alpha}u_{R}^{\beta})(\bar{c}_{R}^{\beta}u_{L}^{\alpha})\right]\,,
\end{equation}
obtaining $C_{1}^{sd,cu}=C_{LL}^{sd,cu}$ and $C_{5}^{sd,cu}=-2C_{LR}^{sd,cu}$.

After integrating out the gauge bosons, we also obtain the following
effective operators contributing to charged lepton flavour-violating
processes,
\begin{flalign}
\mathcal{L}_{\mathrm{EFT}}\supset & -\frac{4G_{F}}{\sqrt{2}}\left[m_{\beta}\left[C_{e\gamma}\right]^{\alpha\beta}\bar{e}_{L}^{\alpha}\sigma^{\mu\nu}e_{R}^{\beta}F_{\mu\nu}+m_{\beta}\left[C_{e\gamma}\right]^{\beta\alpha}\bar{e}_{L}^{\beta}\sigma^{\mu\nu}e_{R}^{\alpha}F_{\mu\nu}\right.\\
 & +\left[C_{ee}^{V,LL}\right]^{\alpha\beta\alpha\alpha}(\bar{e}_{L}^{\alpha}\gamma_{\mu}e_{L}^{\beta})(\bar{e}_{L}^{\alpha}\gamma^{\mu}e_{L}^{\alpha})+\left[C_{ee}^{V,LR}\right]^{\alpha\beta\alpha\alpha}(\bar{e}_{L}^{\alpha}\gamma_{\mu}e_{L}^{\beta})(\bar{e}_{R}^{\alpha}\gamma^{\mu}e_{R}^{\alpha})\nonumber \\
 & +\left[C_{ee}^{V,LR}\right]^{\alpha\alpha\alpha\beta}(\bar{e}_{L}^{\alpha}\gamma_{\mu}e_{L}^{\alpha})(\bar{e}_{R}^{\alpha}\gamma^{\mu}e_{R}^{\beta})+\left[C_{ee}^{V,RR}\right]^{\alpha\beta\alpha\alpha}(\bar{e}_{R}^{\alpha}\gamma_{\mu}e_{R}^{\beta})(\bar{e}_{R}^{\alpha}\gamma^{\mu}e_{R}^{\alpha})+\mathrm{h.c.} \, , \nonumber 
\end{flalign}
where all Wilson coefficients are dimensionless, including those of
the dipole operators, and where $\alpha,\beta=e,\mu,\tau$ and $\alpha\neq\beta$.
Our models generates the four-lepton structures at tree-level with
the following Wilson coefficients
\begin{flalign}
\left[C_{ee}^{V,LL}\right]^{\alpha\beta\alpha\alpha} & =\eta_{\mathrm{RGE}}\left(\frac{\sqrt{2}}{4G_{F}}\right)\left[\frac{\hat{\kappa}_{\alpha\beta}^{e_{L}}\hat{\kappa}_{\alpha\alpha}^{e_{L}}}{2M_{Z'_{12}}^{2}}+\frac{\hat{\xi}_{\alpha\beta}^{e_{L}}\hat{\xi}_{\alpha\alpha}^{e_{L}}}{2M_{Z'_{23}}^{2}}+\frac{\hat{\omega}_{\alpha\beta}^{e_{L}}\hat{\omega}_{\alpha\alpha}^{e_{L}}}{2M_{Z}^{2}}\right]\,,\\
\left[C_{ee}^{V,RR}\right]^{\alpha\beta\alpha\alpha} & =\eta_{\mathrm{RGE}}\left(\frac{\sqrt{2}}{4G_{F}}\right)\left[\frac{\hat{\kappa}_{\alpha\beta}^{e_{R}}\hat{\kappa}_{\alpha\alpha}^{e_{R}}}{2M_{Z'_{12}}^{2}}+\frac{\hat{\xi}_{\alpha\beta}^{e_{R}}\hat{\xi}_{\alpha\alpha}^{e_{R}}}{2M_{Z'_{23}}^{2}}\right]\,,\\
\left[C_{ee}^{V,LR}\right]^{\alpha\beta\alpha\alpha} & =\eta_{\mathrm{RGE}}\left(\frac{\sqrt{2}}{4G_{F}}\right)\left[\frac{\hat{\kappa}_{\alpha\beta}^{e_{L}}\hat{\kappa}_{\alpha\alpha}^{e_{R}}}{M_{Z'_{12}}^{2}}+\frac{\hat{\xi}_{\alpha\beta}^{e_{L}}\hat{\xi}_{\alpha\alpha}^{e_{R}}}{M_{Z'_{23}}^{2}}\right]\,,\\
\left[C_{ee}^{V,LR}\right]^{\alpha\alpha\alpha\beta} & =\eta_{\mathrm{RGE}}\left(\frac{\sqrt{2}}{4G_{F}}\right)\left[\frac{\hat{\kappa}_{\alpha\alpha}^{e_{L}}\hat{\kappa}_{\alpha\beta}^{e_{R}}}{M_{Z'_{12}}^{2}}+\frac{\hat{\xi}_{\alpha\alpha}^{e_{L}}\hat{\xi}_{\alpha\beta}^{e_{R}}}{M_{Z'_{23}}^{2}}\right]\,,
\end{flalign}
where the $\eta_{\mathrm{RGE}}$ coefficients encode RGE effects from
running from the high scale $M_{Z'_{12}}\sim10^{3}$ TeV or
$M_{Z'_{23}}\sim10$ TeV down to 1 GeV. By using {\tt DsixTools}~\cite{Celis:2017hod,Fuentes-Martin:2020zaz} we typically find $\eta_{\mathrm{RGE}}\sim0.8-0.9$ for the leptonic
operators studied here. In contrast, the dipole operators are generated
at 1-loop, with the heavy vector-like fermions also running in the
loop, see Fig.~\ref{fig:diagram_mu_e_gamma}. For example, in case of $\mu\to e\gamma$ contributions, from $Z'_{12}$ exchange we obtain
\begin{equation}
\left[C_{e\gamma}\right]^{\alpha\beta}=\left(\frac{\sqrt{2}}{4G_{F}}\right)\frac{\eta_{\mathrm{RGE}}}{32\pi^{2}M_{Z_{12}'}^{2}}\sum_{a=1}^{6}\left[\hat{\kappa}_{\beta a}^{e_{L}}\hat{\kappa}_{\alpha a}^{e_{L}}F\left(\frac{M^{2}_{a}}{M^{2}_{Z'_{12}}}\right)+\frac{M_{a}}{m_{\beta}}\hat{\kappa}_{\beta a}^{e_{R}}\hat{\kappa}_{\alpha a}^{e_{L}}G\left(\frac{M^{2}_{a}}{M^{2}_{Z'_{12}}}\right)\right]\,,
\end{equation}
\begin{equation}
\left[C_{e\gamma}\right]^{\beta\alpha}=\left(\frac{\sqrt{2}}{4G_{F}}\right)\frac{\eta_{\mathrm{RGE}}}{32\pi^{2}M_{Z_{12}'}^{2}}\sum_{a=1}^{6}\left[\hat{\kappa}_{\beta a}^{e_{R}}\hat{\kappa}_{\alpha a}^{e_{R}}F\left(\frac{M^{2}_{a}}{M^{2}_{Z'_{12}}}\right)+\frac{M_{a}}{m_{\beta}}\hat{\kappa}_{\beta a}^{e_{L}}\hat{\kappa}_{\alpha a}^{e_{R}}G\left(\frac{M^{2}_{a}}{M^{2}_{Z'_{12}}}\right)\right]\,,
\end{equation}
and similarly for $Z'_{23}$ by replacing $M_{Z'_{12}}\rightarrow M_{Z'_{23}}$
and $\kappa\rightarrow\xi$. The loop functions $F(x)$ and
$G(x)$ are
\begin{equation}
F(x)=\frac{5x^{4}-14x^{3}+39x^{2}-38x-18x^{2}\log x+8}{12(1-x)^{4}}\,,\quad G(x)=\frac{x^{3}+3x-6x\log x-4}{2(1-x)^{3}}\,.
\end{equation}
Then finally one can compute the branching fractions of $e_{\beta}\rightarrow e_{\alpha}\gamma$
and $e_{\beta}\rightarrow3 \, e_{\alpha}$ as \cite{Kuno:1999jp}
\begin{equation}
\mathcal{B}(e_{\beta}\rightarrow e_{\alpha}\gamma)=\left(\frac{4G_{F}}{\sqrt{2}}\right)^{2}\alpha_{\mathrm{QED}} \, m_{\beta}^{5}\tau_{\beta}\left(\left|\left[C_{e\gamma}\right]^{\alpha\beta}\right|^{2}+\left|\left[C_{e\gamma}\right]^{\beta\alpha}\right|^{2}\right)\,,
\end{equation}
\begin{flalign}
\mathcal{B}(e_{\beta}\rightarrow3 \, e_{\alpha})= & 2\left(\left|\left[C_{ee}^{V,LL}\right]^{\alpha\beta\alpha\alpha}\right|^{2}+\left|\left[C_{ee}^{V,RR}\right]^{\alpha\beta\alpha\alpha}\right|^{2}\right)+\left(\left|\left[C_{ee}^{V,LR}\right]^{\alpha\beta\alpha\alpha}\right|^{2}+\left|\left[C_{ee}^{V,LR}\right]^{\alpha\alpha\alpha\beta}\right|^{2}\right) \nonumber \\
 & +32\left[\mathrm{log}\left(\frac{m_{\beta}^{2}}{m_{\alpha}^{2}}\right)-\frac{11}{4}\right]\left(\left|e\left[C_{e\gamma}\right]^{\alpha\beta}\right|^{2}+\left|e\left[C_{e\gamma}\right]^{\beta\alpha}\right|^{2}\right) \nonumber \\
 & +\mathrm{Re}\left[e\left[C_{e\gamma}\right]^{\alpha\beta}\left(\left[C_{ee}^{V,LL}\right]^{\alpha\beta\alpha\alpha}\right)^{*}\right]+\mathrm{Re}\left[e\left[C_{e\gamma}\right]^{\beta\alpha}\left(\left[C_{ee}^{V,RR}\right]^{\alpha\beta\alpha\alpha}\right)^{*}\right]\,,
\end{flalign}
where $e=\sqrt{4\pi\alpha_{\mathrm{QED}}}$ is the QED gauge coupling.

\providecommand{\href}[2]{#2}\begingroup\raggedright\endgroup

\end{document}